\documentclass{vgtc}                          % final (conference style)
\ifpdf%                                % if we use pdflatex
  \pdfoutput=1\relax                   % create PDFs from pdfLaTeX
  \usepackage{graphicx}                % allow us to embed graphics files
  \DeclareGraphicsExtensions{.pdf,.png,.jpg,.jpeg} % for pdflatex we expect .pdf, .png, or .jpg files
\else%                                 % else we use pure latex
  \usepackage{graphicx}                % allow us to embed graphics files
  \DeclareGraphicsExtensions{.eps}     % for pure latex we expect eps files
\fi%

\graphicspath{{figures/}{pictures/}{images/}{./}} % where to search for the images

\usepackage{microtype}                 % use micro-typography (slightly more compact, better to read)
\PassOptionsToPackage{warn}{textcomp}  % to address font issues with \textrightarrow
\usepackage{textcomp}                  % use better special symbols
\usepackage{mathptmx}                  % use matching math font
\usepackage{times}                     % we use Times as the main font
\usepackage{cite}                      % needed to automatically sort the references
\usepackage{tabu}                      % only used for the table example
\usepackage{booktabs}                  % only used for the table example
\usepackage{balance}       % to better equalize the last page
\usepackage{graphics}      % for EPS, load graphicx instead 
\usepackage[T1]{fontenc}   % for umlauts and other diaeresis
\usepackage{txfonts}
\usepackage{mathptmx}
\usepackage[pdflang={en-US},pdftex]{hyperref}
\usepackage{color}
\usepackage{booktabs}
\usepackage{textcomp}
\usepackage{enumitem}
\usepackage{todonotes}
\usepackage{tabularx, booktabs} %% Load packages that you use
\usepackage{graphicx} 
\usepackage{adjustbox}

\graphicspath{{figures/}{pictures/}{images/}{./}} % where to search for the images

\onlineid{0}

\vgtccategory{Research}

\vgtcinsertpkg

\title{DeepVA: Bridging Cognition and Computation through Semantic Interaction and Deep Learning}

\author{
Yali Bian\thanks{e-mail: yali@vt.edu}\\ %
        \scriptsize Virginia Tech %
\and John Wenskovitch\thanks{e-mail:jw87@.edu}\\ %
     \scriptsize Virginia Tech %
\and Chris North\thanks{e-mail:north@vt.edu}\\ %
     \scriptsize Virginia Tech %  
     }

\teaser{
    \centering
    \includegraphics[width=0.824\linewidth]{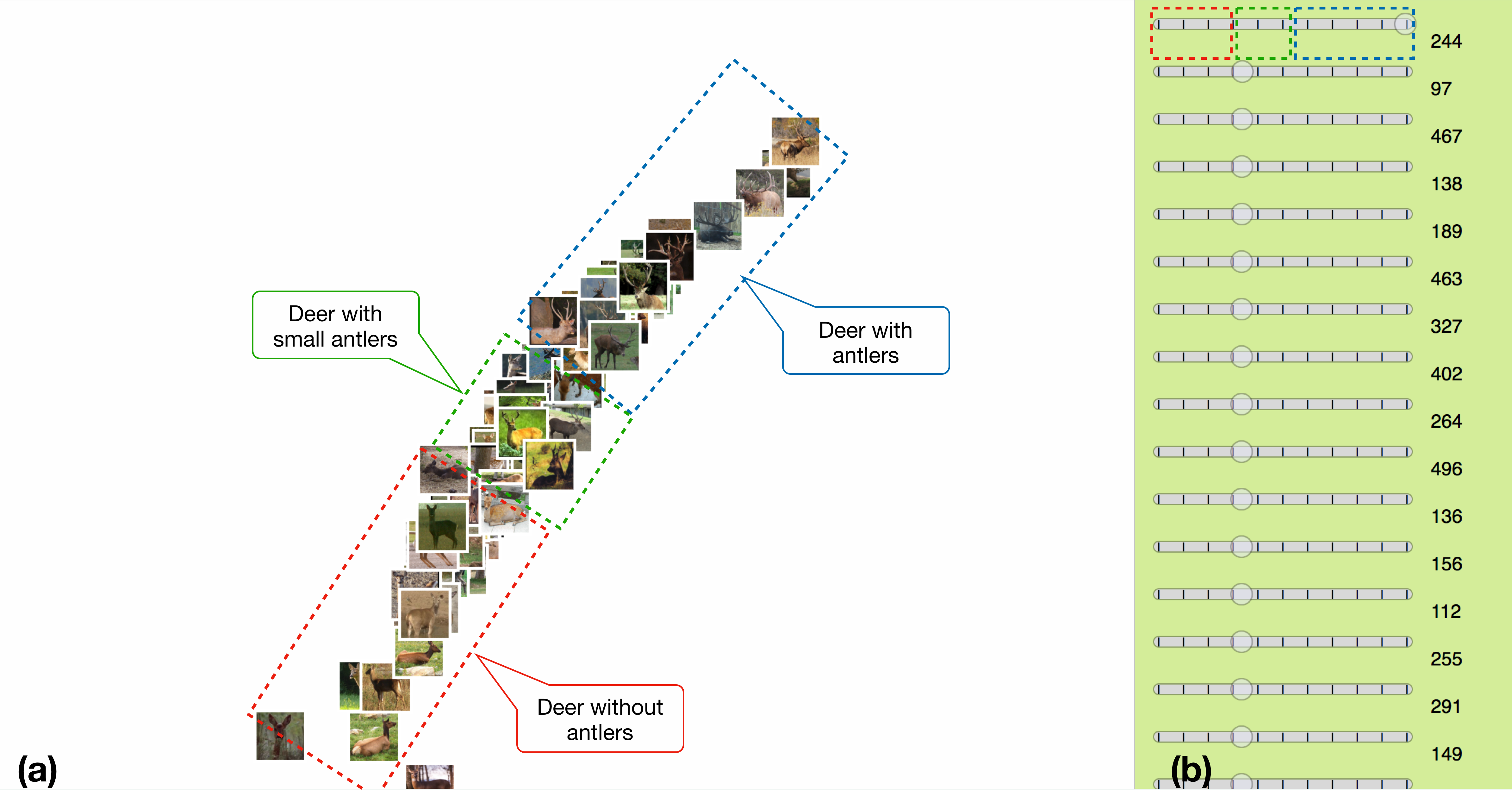}
    \caption{
        DeepVA projection of deer images based on the concept ``antler.''
        (a) Images are projected using features in their deep learning representations.
        (b) The weights of features used in the projection.
        DeepVA couples cognition and computation through the use of semantic interaction on high-level deep learning features. 
        In this case, DeepVA captures a user's organizing intent about the high-level visual concept ``deer with antlers,'' and maps it to the deep learning feature~$d_{244}$ that contains relevant semantic information.
    }
    \label{fig:vis}
}

\abstract{

This paper examines how \textit{deep learning} (DL) representations, in contrast to traditional engineered features, can support \textit{semantic interaction} (SI) in visual analytics.  SI attempts to model user's cognitive reasoning via their interaction with data items, based on the data features.
We hypothesize that DL representations contain meaningful high-level abstractions that can better capture users' high-level cognitive intent.
To bridge the gap between cognition and computation in visual analytics, 
we propose \textit{DeepVA} (Deep Visual Analytics), which uses high-level deep learning representations for semantic interaction instead of low-level hand-crafted data features.
To evaluate DeepVA and compare to SI models with lower-level features, we design and implement a system that extends a traditional SI pipeline with features at three different levels of abstraction. 
To test the relationship between task abstraction and feature abstraction in SI, we perform visual concept learning tasks at three different task abstraction levels, using semantic interaction with three different feature abstraction levels. 
DeepVA effectively hastened interactive convergence
between cognitive understanding  and  computational modeling of the data, especially in high abstraction tasks.
}

\begin{document}
\firstsection{Introduction}
\label{sec:introduction}
\maketitle
Coupling cognition and computation through interactivity is a challenging and important research topic of visual analytics (VA)~\cite{cook2005illuminating}.
\textit{Semantic Interaction~(SI)}~\cite{7160906} seeks to enable coupling during sensemaking~\cite{pirolli_2005} tasks.
To synthesize information, users can manipulate and organize data items to express their reasoning.
Analytic methods can then help the synthesis process by systematically observing these user interactions and learning a synthesis model.
Systems can respond to the user's cognitive reasoning process by mapping the interactive intents to underlying model parameters on data features~\cite{Endert:ji}.

However, a major problem is that it can be difficult to capture users' semantic intents in complex sensemaking tasks because of the gap between high-level cognition and low-level computation.
For example, Endert's study~\cite{Endert:he} revealed a distinction between high-level cognitive features and low-level data features.
Since it typically requires significant cognitive effort to reduce high-level concepts to low-level features through the course of incremental  formalization~\cite{shipman1994supporting}, 
SI models attempt to relieve this requirement by indirectly learning the mapping.
Analytic models must transform users' high-level cognitive concepts or topics into low-level data features. 
But, the low-level data features might not support a clear mapping, making it very difficult to efficiently capture subtle high-level intents.

Simultaneously, recent research on deep learning presents promising solutions to address the challenge of bridging the semantic gap.
Deep learning can automatically learn meaningful data features~(representations) from a large amount of data, and it outperforms many traditional machine learning techniques in many fields~\cite{LeCun:2015dt}.
For example, \textit{Convolutional Neural Networks} (CNN) have been very successful for image recognition tasks~\cite{krizhevsky2012imagenet}.
The DL approach has several benefits: it relieves the need to manually engineer good features for the underlying machine learning model; 
the meaningful representations can better capture high-level concepts needed by the analyst~\cite{kim2017interpretability};
and, the trained DL model can often be transferred to other applications using similar datasets~\cite{razavian2014cnn}.

Therefore, to overcome the semantic gap challenge, this paper explores the potential use of deep learning features for SI to support visual analytics for interactive data synthesis tasks.
We hypothesize that
%\begin{enumerate}[noitemsep]
%\item 
deep learning representations can capture the meaning of users subtle interactive intents in SI. Thus it enables more complex and effective interactive syntheses of the data.
%\item 
\textbf{Specifically, in this paper, we investigate how deep learning representations can improve  learning from users' \textit{observation-level interactions (OLI)}~\cite{Endert:ji}, a particular type of semantic interaction that involves organizing data points in a dimension-reduced projection.} % of high-dimensional data. 
%\end{enumerate}
To generalize the effect, we hypothesize that higher abstraction-level features will better support higher abstraction-level tasks with SI.

To explore this hypothesis, we design a system that focuses on providing analysts with OLI interactions for image sorting and spatial organizing.
We extend the SI pipeline model in our designed system %, representing human cognition with automatically learned DL features, 
by introducing  data features at different abstraction levels into the SI process.
We examine image feature sets at three different abstraction levels: color histogram (low-level), SIFT features (mid-level), and deep learning representations (high-level).  
%We compare the high-level deep learning representations with the other two lower-level engineered feature sets in SI models.
%We seek to better understand the effects of the deep learning representations in improving VA systems with SI. 
To make a systematic assessment, we perform a 3$\times$3 case study.
SI with high-level DL features ($SI_{high}$, also denoted as DeepVA) is compared to SI with the lower-level engineered feature sets ($SI_{low}$ and $SI_{mid}$).
%SI is tested with three types of features: the learned features are compared against two lower-level engineered feature sets.
For each, a synthesis task is performed at three different abstraction levels of visual concepts: ``ground'' (low-level), ``deer'' (mid-level), and ``deer with antler'' (high-level).

The results demonstrate initial %proof-of-existence
evidence for the validity of our hypotheses. 
SI with DL features (DeepVA) effectively and efficiently supported synthesis tasks at all three task abstraction levels.
Because DL representations contain relatively high-level concepts, users' complex high-level intents were more directly mapped to these features with less interactive cost compared to the lower-level feature sets.
In other words, DeepVA better optimized the coupling process between cognition and computation. 

The contributions of this paper are: 
\begin{enumerate}[noitemsep]
    \item A new SI model, DeepVA, that uses deep learning techniques to enhance VA systems with SI in solving more difficult synthesis tasks.
    
    \item An extended SI system designed for image analysis that provides the evaluation and comparison between SI models with features in different abstraction levels. 
    
    \item A systematic evaluation of nine case studies is performed to examine DeepVA's ability to address exploratory synthesis tasks, in comparison to the use of lower-level data features. 
\end{enumerate}

\section{Related Work}

This section covers many related VA and DL techniques that have influenced DeepVA. 
First, we discuss VA with SI in details.
After this, we describe deep learning and related visual analytic techniques.

\subsection{Visual Analytics with SI} 
\label{sec-VA}
Instead of directly interacting with difficult-to-understand parameters of underlying models, semantic interaction (SI)~\cite{Endert:he} makes use of natural interactions within the projection to learn the intent of the analyst, which is then used to update underlying models.
Concepts, which would be difficult to externalize and pass into underlying models through parameters, can be more easily expressed.
It is the system's responsibility to perform the communication between cognition and computation through the 2D visual layout.

Several machine learning models have been explored to solve the bi-directional transforms. OLI (Observation-Level Interaction)~\cite{Endert:ji} and it generalized frameworks, 
V2PI~(Visual to Parametric Interaction)~\cite{Leman:2013it} and Bayesian visual analytics~(BaVA)~\cite{house2015bayesian}, show how popular dimension reduction models, such as~\textit{Weighted Multidimensional Scaling} (WMDS)~\cite{schiffman1981introduction}, can be inverted to allow the direct manipulation of the points within the projection.
Semantic interaction has been applied to text analytics in ForceSPIRE~\cite{endert2012semantic}, and high-dimensional quantitative data in Andromeda~\cite{Zeitz:2018:BIV:3144687.3144715}. 
Semantic interaction was also applied on images in ACTIVECANVAS~\cite{DBLP:journals/corr/HodasE16}, to add extra dimensions or attributes to the data based on the domain expertise of the user expressed via user interactions.
To steer nonlinear data models for users' complex, high-level domain knowledge, AxiSketcher~\cite{7534876} introduces drawing as an interaction to express their complex intents, and nonlinear axis mapping methods to model the intents.

In this paper, deep learning representations are used as data features.
For these difficult-to-understand features, semantic interaction is more appropriate than direct manipulation.
We apply semantic interactions to the learned representations, instead of to the basic data features that were used in the above systems.
While both our paper and ACTIVECANVAS enable analysis of image data, our work focuses on mapping interactions to learned features to boost human sensemaking activities, rather than using the performed interactions as features for future processing.

Our work in this study has much in common with AxiSketcher: both try to capture users' complex intent and adapt it to underlying models through SI.
However, they differ in how users intents are modeled.
In DeepVA, complex intents are inferred and modeled linearly to high-level meaningful representations.
While in AxiSketcher, the non-linear model is used to represent intents with low-level features.

\subsection{Deep Learning Overview}
The ability to extract abstract but meaningful features makes deep learning an attractive approach for  complex problems like visual object recognition and natural language understanding.

\subsubsection{Representation as High-Level Features}
\label{sec:representation}

In contrast to conventional machine learning techniques, deep learning can extract useful and meaningful representations in complex data automatically through the usage of deep artificial neural networks.
As stated by LeCun et al., ``Deep learning allows computational models that are composed of multiple processing layers to learn representations of data with multiple levels of abstraction"~\cite{LeCun:2015dt}.
In deep neural networks, initial layers are used to learn lower-level concepts, such as local features in images. 
Representations in later layers are computed based on the previous layers and contain higher-level concepts that result from the combination of these earlier layers.
For instance, when using CNN on images, local combinations of edges form motifs, which then assemble into parts, which assemble into objects ~\cite{krizhevsky2012imagenet}. 

Deep learning representations are typically used internally to the learning process to support the final layer for analytics tasks such as classification.
However, as high-level abstract features, deep learning representations are powerful and could be used outside of deep learning processes to solve other problems.
Razavian et al.~\cite{Razavian:wa} conducted experiments for different recognition tasks based on traditional classification methods and CNN representations as input features.
Their results show that features obtained from deep learning with convolutional nets should be the primary candidate in most visual recognition tasks.
Athiwaratkun et al.~\cite{athiwaratkun2015feature} showed that even pre-trained CNN is much more useful as generic feature extractors for computer vision tasks, than commonly used engineered features, such as \textit{Scale-Invariant Feature Transform} (SIFT)~\cite{Lowe2004}.
Amir et al.~\cite{taskonomy2018} provides a fully computational approach for modeling the structure of space of visual tasks for images through transfer learning techniques. 
Been et al.~\cite{kim2017interpretability} introduces Concept Activation Vectors (CAVs), which provide an interpretation of a neural net's internal state in terms of human-friendly concepts.
% These research results are indicating the possibilities of the connections between complex concepts and learned representations for our DeepVA.

\subsubsection{Visual Analytics for Deep Learning}
Deep learning representations and structures are difficult to interpret because of the innate complexity and nonlinear structure of deep neural networks, leading to the need for interpretable or explainable methods.
Fred et. al.~\cite{hohman2018visual} summarizes thoroughly the state-of-the-art of visual analytics in deep learning research for model explanation, interpretation, debugging, and improvement. 
Jaegul et. al.~\cite{8402187}, reviews existing visual analytics techniques in making deep learning interpretable and controllable by humans from the perspective of visual analytics, information visualization, and machine learning.
To help researchers comprehend the abstract representations or complicated structures of deep learning, visual analytics systems have  been developed~\cite{7536654,Smilkov:2017to,Wongsuphasawat:bb}. Beyond structures, other related work that helps researchers understand the dimensions of deep learning representations has been well described~\cite{Girshick:vu, salimans2017pixelcnn++, Zeiler:2014fr}.

As a side benefit, DeepVA offers a novel solution to the problem of interpreting deep learning models.
Our method emphasizes an interactive approach that exploits users' data domain knowledge to enable discovery of the semantics of mysterious deep learning features.

\subsubsection{Deep Learning for Visual Analytics}
Recently, there are visual analytics (VA) approaches integrating Deep learning algorithms to assist users in data analysis. 
Hsueh-Chien et. al.~\cite{8265023} uses CNN techniques to assist users in volume visualization designing through facilitating user interactions with high-dimensional features with deep-learning methods.
Yi et. al.~\cite{WANG201766} develops interactive deep learning method for segmenting moving objects, that only small number of user interventions are needed to provide results sufficiently accurate to be used as ground truth.
Sharkzor~\cite{DBLP:journals/corr/abs-1802-05316} is a web application for deep learning assisted image sort and summary. 
Sharkzor is perhaps most conceptually similar to DeepVA, and leverages deep learning algorithms  to learn an image classifier based on the user's previous image clustering activity.
While the technique is related, our approach emphasizes how deep learning representations help users organize a 2D space (dimension reduction) while maintaining a higher level of cognitive state and fostering the incremental sensemaking process from the OLI perspective.

\section{System Design and Implementation}

\begin{figure}
\centering
  \includegraphics[width=1.0\columnwidth]{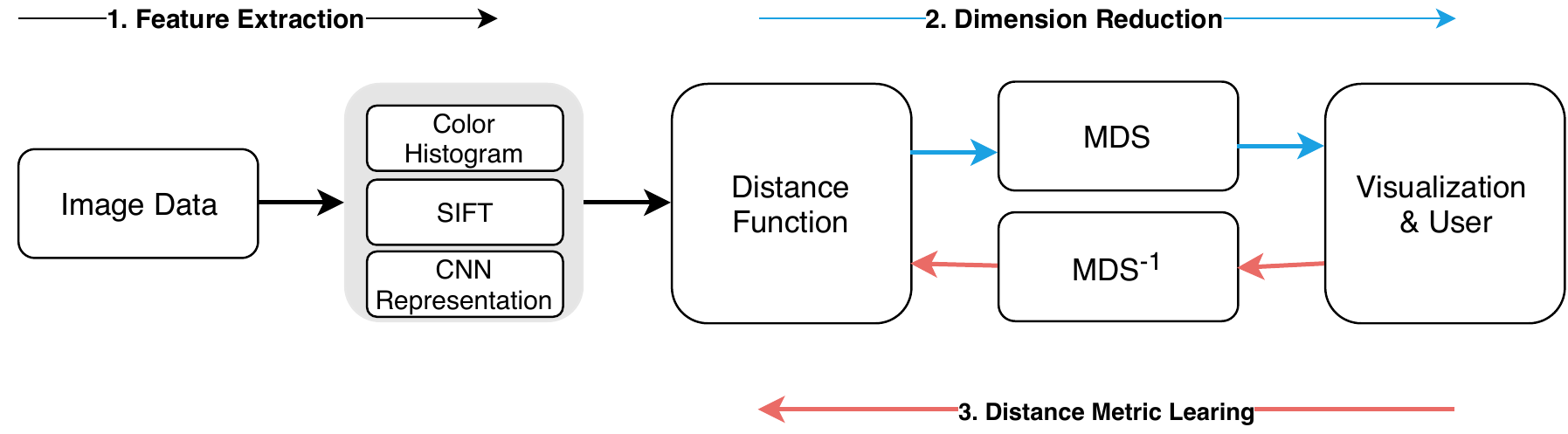}
  \caption{The system pipeline is composed of three important components.
  %to evaluate and compare the effectiveness of deep learning with other lower-level features in support SI on sensemaking tasks:
  1. Feature extraction: the system extends the SI pipeline with  features at three different abstraction levels. For image data: $F_{low}$ is the color histogram, $F_{mid}$ is the SIFT feature, and $F_{high}$ is the DL representation;
  2. Dimension reduction: projects the data based on the learned distance function;
  3. Distance metric learning: updates the distance function based on the analyst's interactions in the visualization.  
  }
  ~\label{fig:deepva}
\end{figure}

To explore and evaluate the effectiveness of different abstraction levels of data features in semantic interactions (OLI in particular) on capturing user cognition, we design a system prototype  (Fig.~\ref{fig:deepva}) based on Andromeda~\cite{Zeitz:2018:BIV:3144687.3144715}. We focus on the process of feature extraction, as well as the other two important components in OLI models (dimension reduction and distance metric learning). 

In this paper, the system is designed and implemented for image analysis tasks. 
However, the model generalizes to tasks, such as document analysis, or numerical data analysis. OLI systems are designed to enable users to create customized projections of high-dimensional data that represent their own expertise. Users express hypothesized similarities by directly manipulating a subset of points in the display. Semi-supervised metric learning then learns a new weighted distance function based on the data features, and updates the projection accordingly.  Typically, the user's goal is to create a spatial projection that captures a desired cognitive concept. In the case of image data, that means organizing images according to important content within the images. For example, an analyst might spatially organize images of animals according to similarities in a biological taxonomy,  the physical layout of a zoo, or their own preconceptions about animal similarities.  

The main challenge of OLI systems is whether the metric learning can adequately capture the user's intended structure given the available data features. Our hypothesis is that higher-level features, such as deep learning representations, will be better able to capture user's high-level OLI intents.

\subsection{Feature Extraction}
To examine the influence of features at different abstraction levels on OLI models, %especially the high-level deep learning features, compared with traditional hand-craft features, 
we focus on exploratory analysis of image data with three commonly used features: color histogram ($\mathbf{F_{low}}$), the Scale-Invariant Feature Transform (SIFT) ($\mathbf{F_{mid}}$), and CNN representation ($\mathbf{F_{high}}$).  The first two represent traditional low-level engineered feature sets, while the latter represents high-level learned features.
As shown in Fig.~\ref{fig:deepva}, the system with the  three different levels of images features are equivalently used as input features in the SI system. 
To improve the modeling efficiency and remove unwanted variation, the number of features is limited to~512 in all three feature extraction methods.

$\mathbf{F_{low}}$~\textbf{- Color Histogram: }
The color histogram~\cite{391417} is a representation of the distribution of colors in an image. 
Color histograms contain only basic low-level pixel information, without any semantic information. 
Images with an RGB color model were transferred to a color histogram that contains $255\times3$ dimensions. 
To reduce the feature size to \textit{512D}, a \textit{Principal Component Analysis} (PCA)~\cite{WOLD198737} dimension reeducation model was performed on the  histogram.

$\mathbf{F_{mid}}$~\textbf{- SIFT: }
SIFT has been the most widely-used hand-crafted feature generator for content-based image retrieval  in the past decade.
Unlike CNN, SIFT only describes local patterns as features~\cite{Lowe2004} in images and only local semantic information is embedded in each dimension.
We use OpenCV~\cite{itseez2015opencv}, a widely-used computer vision library, to extract SIFT keypoints and  descriptors for each image.
However, the features extracted through SIFT are variable size. 
To fix the number of features to \textit{512D}, \textit{Bag of Visual Words} (BOVW)~\cite{yang2007evaluating} is used to select the most representative 512 SIFT features for the images. 

$\mathbf{F_{high}}$~\textbf{- CNN Representation: }
Due to the training requirements of deep neural networks, a huge amount of labels are needed, which is difficult for traditional VA models to supply.
Also, deep learning representations are mysterious, analysts could not directly map their intents to representation space.
To address these challenges, we  applied  basic transfer learning in the feature extraction process for CNN representations (Fig.~\ref{fig:deepva}). We use a pre-trained deep learning model as a fixed feature extractor. We freeze the weights for  the entire network except for the final fully connected layer. This last fully connected layer is replaced with a new one with random weights and only this layer is trained by the default statistic model used in OLI models.
We use the pre-trained ResNet (ResNet-18 model)~\cite{he2016deep} on ImageNet~\cite{Krizhevsky:2012:ICD:2999134.2999257}  (removing the last fully-connected layer) as a feature extractor, a CNN model in the torchvision package, which contains commonly used datasets and model architectures for computer vision, on top of the deep learning platform PyTorch~\cite{paszke2017automatic}.
Other deep learning frameworks, including TensorFlow~\cite{abadi2016tensorflow}, Caffe~\cite{jia2014caffe}, and Theano~\cite{bergstra2010theano}, also contain pre-trained deep learning models and could be used to extract DL representations.
The  high-dimensional representations were compressed to~$512D$, using representation transformations in the torchvision package.

\subsection{Dimension Reduction}
OLI systems use dimension reduction models to provide analysts an interactive space and visual feedback about the learned concepts. 
% In this system, the distance function is applied to learn users' cognition with the extracted feature. 
% The distance function could provides the distances between images based on analysts' preference.
% The distances is then used to display the image scatter plot on the visualization with MDS dimension reduction model.

\subsubsection{Distance function}
Similar to previous OLI systems, we use a weighted  distance function to capture and reflect analysts' preference and understanding of the similarities between images, based on the extracted features, in a 2D visual projection.
Initially, the uniformly weighted Euclidean distance function is used for the default projection:
%assumption of analysts' preference about the distance between two image pair based on current extracted feature: 
\begin{equation}
D_w(x_i, x_j)=\sqrt{\sum_{k\in w} {w_k*(x_{i, k}-x_{j, k})^2}}
\label{equation:dist}
\end{equation}

This distance, described in Equation~\ref{equation:dist}, is the Euclidean distance between the normalized features of image $x_i$ and $x_j$, including a weight $w$ applied to each feature that denotes the importance of that feature to the current projection. 
At system initialization, each of the weights associated with the features in the dataset are set to 1, indicating that each feature has equal contribution to the pairwise distances between images. 
%resting length of each link than any other weight. 
These weights are updated in response to user interaction in the process of learning a new distance function to capture the users' cognitive intents, detailed in the next subsections.

\subsubsection{WMDS} 
As in each interactive loop, the updated distances between images are mapped to a 2D visual projection. 
The mapping from representation to visualization helps users  make sense of the highly-weighted features, verifying their concepts through the updated visualization layout. We use multi-dimensional scaling (MDS)
together with the weighted distance function to create a weighted-MDS (WMDS)  projection of the weighted high-dimensional features into the 2D space~\cite{Zeitz:2018:BIV:3144687.3144715}. In the projection, MDS seeks to minimize the mean squared error between the 2D and nD pairwise distances.

\subsubsection{Visualization} 
With the dimension reduction model, the similarity relationships between images are shown in the scatter plot visualization Workspace view. 
The learned weights of  features are also visualized as sliders in the Feature view.
The Workspace View provides a space to perform image movement interactions (OLI), affording the user with the ability to express semantic concepts.
The Feature View displays weights that have been learned for data features.

\textbf{Workspace view}
As shown in Fig.~\ref{fig:vis}a, the Workspace View is a scatterplot of images. %Here we use a dataset of images.
The distance between images reflects their relative similarity as weighted by the  underlying analytic models. 
At first, images scatter in the Workspace based on an equally-weighted dimension reduction of the high-dimensional features.
If two images are positioned close to each other in this initial projection, it implies that these two images are likely similar based on all data features.
There are three interactions provided in the Workspace  for users to explore and view the images. 
Users can drag images to modify the projection and trigger the learning of new weights.
When a user hovers on an image, the values for each feature dimension will be displayed in the Feature View with a yellow circle glyph, and the hovered image will be highlighted with a yellow border.
Clicking a single image or drawing a box around multiple images, causes the images to be highlighted with blue borders, and also displays the associated feature values with blue circles on the Feature view sliders.

\textbf{Feature View}
This view (Fig.~\ref{fig:vis}b) displays the weighted data features.
For  DeepVA, DL representations extracted from transfer learning are used as input features.
These features are visualized by interactive sliders.
The  weight applied to a feature is mapped to the knob position of its slider.
The weights can be directly modified by dragging the knobs.
Updated weights will be automatically applied to update the image projection in the Workspace View.
Additionally, the weights displayed for each feature will  update after the user drags images to update the projection via OLI.
The purpose of interacting with the sliders is primarily for hypothesis testing, and making sense of abstract features.
If users formulate a hypothesis about one feature, they can increase the weight on that feature to test if the new weight creates a layout in the Workspace View that matches the desired outcome of the hypothesis.
The updated image layout in the Workspace View can  reveal some conceptual meaning about the up-weighted feature.
For example, as shown in Fig.~\ref{fig:vis}, the feature $d_{244}$ of the DL representation is linked to the semantic concept ``antlers,'' because when $d_{244}$ was increased in weight, the updated Workspace shows that images of deer are linearly organized  based on how much antlers the deer have.

\subsection{Distance Metric Learning}

% On the visualization, dragging images allows users to manipulate the spatial layout directly by placing items in locations based on the users' domain knowledge and synthesis goals.
% After the user has externalized their synthesis intent by dragging several images, the user can click the ``update'' button.
With OLI, users' concepts are expressed by dragging  elements closer or farther based on desired similarity. 
As discussed previously,  OLI enables analysts to  synthesize concepts  by organizing data elements in a 2D visual interface.
To capture users' cognition, the underlying model should have the ability to automatically capture, infer and associate users' intents with data features through these OLIs.

\subsubsection{$WMDS^{-1}$: Inverse WMDS}
Commonly, a distance metric learning method is used to update the distance function based on OLI interactions with 2D position information.
The underlying distance metric learning method, as  described in Andromeda~\cite{Zeitz:2018:BIV:3144687.3144715}, is executed to infer the user's organizational intent and to recalculate the layout of all images based on the new spatial positions of the dragged images.
An appropriate metric-learning method for WMDS projections is $WMDS^{-1}$, also know as Inverse WMDS.
$WMDS^{-1}$ uses a similar optimization as WMDS, but instead retrains the distance function by updating the feature weights $w$ according to the new distances expressed by the user.

%Since, the weighted Euclidean distance is used on our system, ($WMDS^{-1}$:) will update the distance function, through updating the feature weights $w$ to minimize the following equation, as an optimization problem: 

% This equation is wrong. See Andromeda paper.
%\begin{equation}
%w_{next}=\arg\min_{w} \sqrt{\sum_{k\in w} %{w_k*(x_{i, k}-x_{j, k})^2}}
%\label{equation:wmds}
%\end{equation}

%The update of weight is based on all the moved data points on the visualization, instead of all the data points. 
Through the interactive loop, the analyst iteratively provides  more feedback to incrementally update the distance function. %, and then provide more preference about the layout to update the distance function again. 
With more feedback, the distance function will more accurately reflect  the analyst's intent.
%With different features, the distance function might be trained differently. 

\section{Case Study Design}

% \chris{need to make the design of the experiment more clear that we are testing 3 levels of abstraction in the features and 3 levels of abstraction in the task.  our hypothesis is that the feature abstraction should match the task abstraction, or that Feature abstraction >= task abstraction level, as in Fig 3.
% need to choose a term/phrase like "abstraction level" and stick with it.
% need to justify why the chosen tasks and feature types are at each low, medium, high abstraction levels.
% need to update the figs with this terminology.
% use Low, Med, High instead of 1,2,3 and Basic,Eng,Learned}
\subsection{Research Questions}

We investigate the following research question:
%
%\begin{itemize}
%    \item $Q1$: 
    Given OLI, how well do different data feature abstraction levels support OLI  in learning different user concept  abstraction levels?
%    
%    \item $Q2$: 
    Specifically, does OLI with high abstract DL representations $SI_{high}$ (DeepVA) improve performance at interactively learning users' higher-level concepts?
%\end{itemize}

We hypothesize that $SI_{high}$ (our proposed method, DeepVA), will facilitate higher-level synthesis tasks better than SI with low-level features (SI with SIFT features $SI_{mid}$, SI with color histogram $SI_{low}$), since it offers a better match between high-level cognitive concepts and high-level computational features. 
Thus, the high-level features should be able to more easily capture the high-level concepts expressed by the user.  
SI with low-level features has been well evaluated for different data types~\cite{Bradel:dd,Endert:he,Zeitz:2018:BIV:3144687.3144715}.
However, there is no study of SI with more abstract learned features. % or engineered features.
To assess this hypothesis, we compare $SI_{high}$ used in DeepVA with two other variants of SI using lower-level feature sets, $SI_{mid}$ and $SI_{low}$. 

\subsection{Design Rationale}

\begin{figure}
\centering
  \includegraphics[width=1.0\columnwidth]{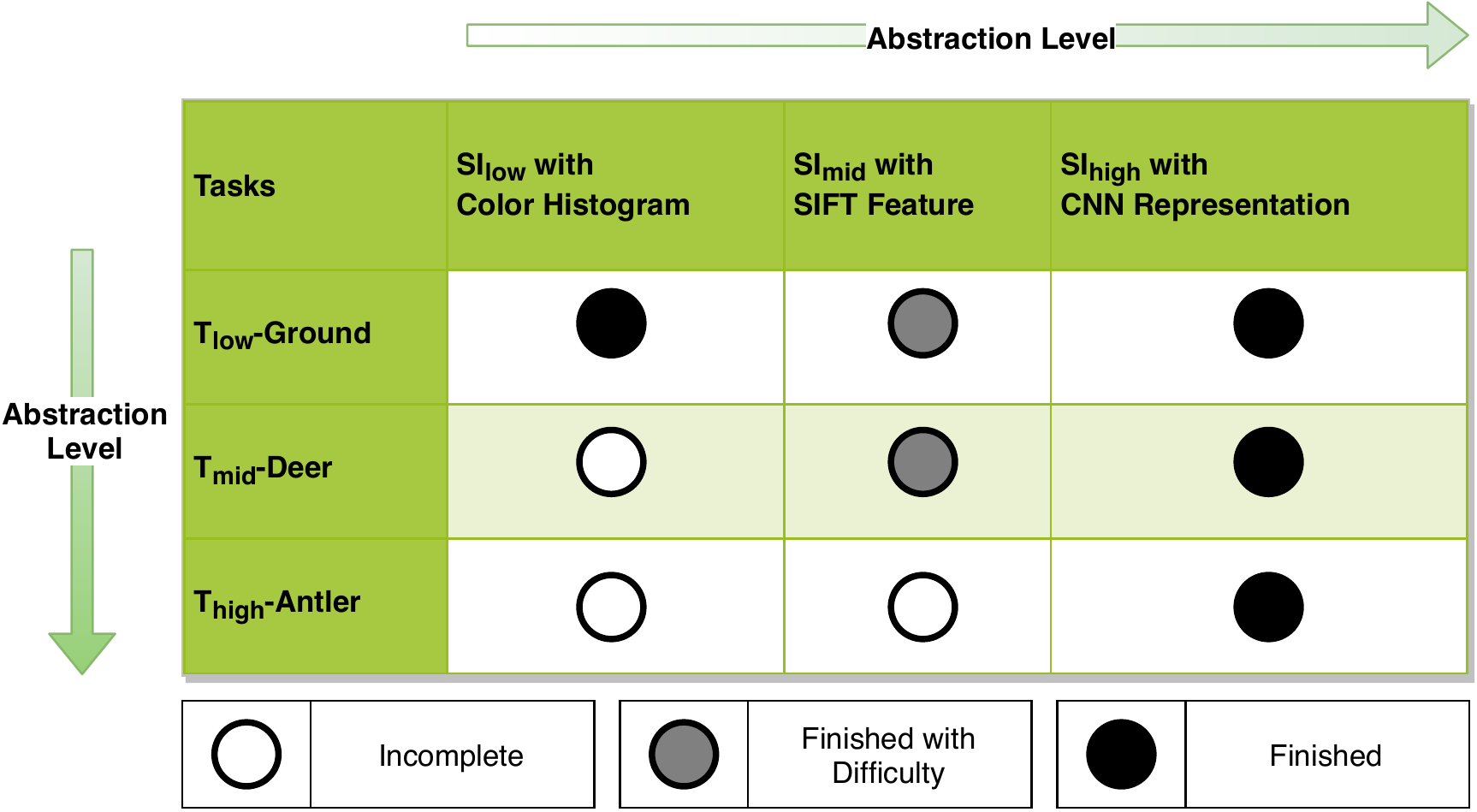}
  \caption{A summary of our study design and results:  concept learning tasks at three different abstraction levels, and  OLI SI methods with features at three different abstraction levels. The results show that SI with higher-level features can accomplish higher-level user tasks (as well as lower-level tasks).}
  ~\label{fig:tasks}
\end{figure}

For a systematic evaluation, we perform 9 case studies on  visual concept learning tasks at three different conceptual abstraction levels, using OLI SI methods with three different feature abstraction levels. 
As shown in Fig.~\ref{fig:tasks}, this $3\times3$ factorial study helps illuminate our research questions.

Visual concept learning tasks have been well explored in the field of computer vision.
There are three commonly used types of image features with different levels of semantic information: Color Histogram (low abstraction feature); SIFT features (mid abstraction feature); and CNN Representation (high abstraction features).
Likewise, visual concepts can be  categorized and ranked according to their level  of abstraction.
For example, the concept ``deer with antlers'' is more high-level than ``deer,'' because ``deer with antlers'' can be defined only when ``deer'' is defined, and requires a more advanced structure of deer images. Whereas, ``ground'' is conceptually more low-level and more easily maps directly to low-level features like ``green''.
We adapt the visual concept learning task to VA, as it is well representative of many types of sensemaking tasks in which VA users  sort and organize images (or other types of information) to synthesize high-level visual concepts.

\subsection{Data}
Studies were conducted on the STL10 dataset~\cite{coates2011analysis}, which   contains 10 classes of objects such as animals and cars.
We choose STL10 over other commonly used datasets, such as CIFAR-10 or MNIST, for two reasons: 
images in STL10 have higher resolution (96x96 pixels), that can help  users and DL representations  to detect subtle visual concepts, such as antlers; 
images in STL10 have a variety of simple scenes, such as ``blue sky'' and ``green ground,'' in addition to complex objects.

\subsection{Task Design}
As shown in Fig.~\ref{fig:tasks}, we designed  visual concept learning tasks at three different difficulty levels.
For each task, with selected images loaded in our system, we attempted to express the desired visual concept by using OLI to drag a subset of the images to organize a representative 2D structure.  For each, we tried interacting with subsets of different sizes, until the underlying model was able to effectively learn the desired concept, which we could  recognize by how well the rest of the images were organized by the model and whether it matched the desired concept.
A better SI feature set should require fewer interactions by the analyst  to express the desired concept to the underlying model.  In general, SI users want to effectively train their models with as few interactions as possible.

$\mathbf{TASK_{low}}$\textbf{-Ground:}~``Ground'' is simple and basic concept that should only need local features to capture well, since it can be directly described by basic color features.
100 images are loaded into visual interface, including 50 images about green ground, and 50 images about blue sky or sea. 

$\mathbf{TASK_{mid}}$\textbf{-Deer:}~``Deer'' is a complex object-level concept that requires many local features to be combined. 
100 images are used, including 50 images about deer, and 50 images about birds. We used images of deer and birds since they have similar backgrounds containing trees, thus not falling into a trap like wolf==snow~\cite{Ribeiro:2016:WIT:2939672.2939778}.
 
$\mathbf{TASK_{high}}$\textbf{-Antler:}~To design an even more complex task, we choose the visual concept ``antler'' that contains more subtle concept refinements, since it has similar shapes as background trees, and also must be connected on the head of a ``deer.''
In this task, 100 images are used, including 50 images about deer with antlers, and 50 without antlers.

\section{Qualitative Results}
\label{sec:case_study} 
We describe in detail three of the case studies about $TASK_{mid}$-Deer, using $SI_{high}$, $SI_{mid}$, and $SI_{low}$.
The other 6 studies about $TASK_{low}$ and $TASK_{high}$ are included in the quantitative evaluation results in Fig.~\ref{fig:tasks} and~\ref{fig:cost}.

\subsection{\texorpdfstring{$\mathbf{TASK_{mid}: SI_{high}}$}~\textbf{with CNN representation}}
In this method, the underlying analytic model  used CNN representations to capture and infer the analyst's goal, ``deer'', based on  OLI interactions.

\begin{figure}[tb]
\centering
{
    \includegraphics[width=0.9\columnwidth]{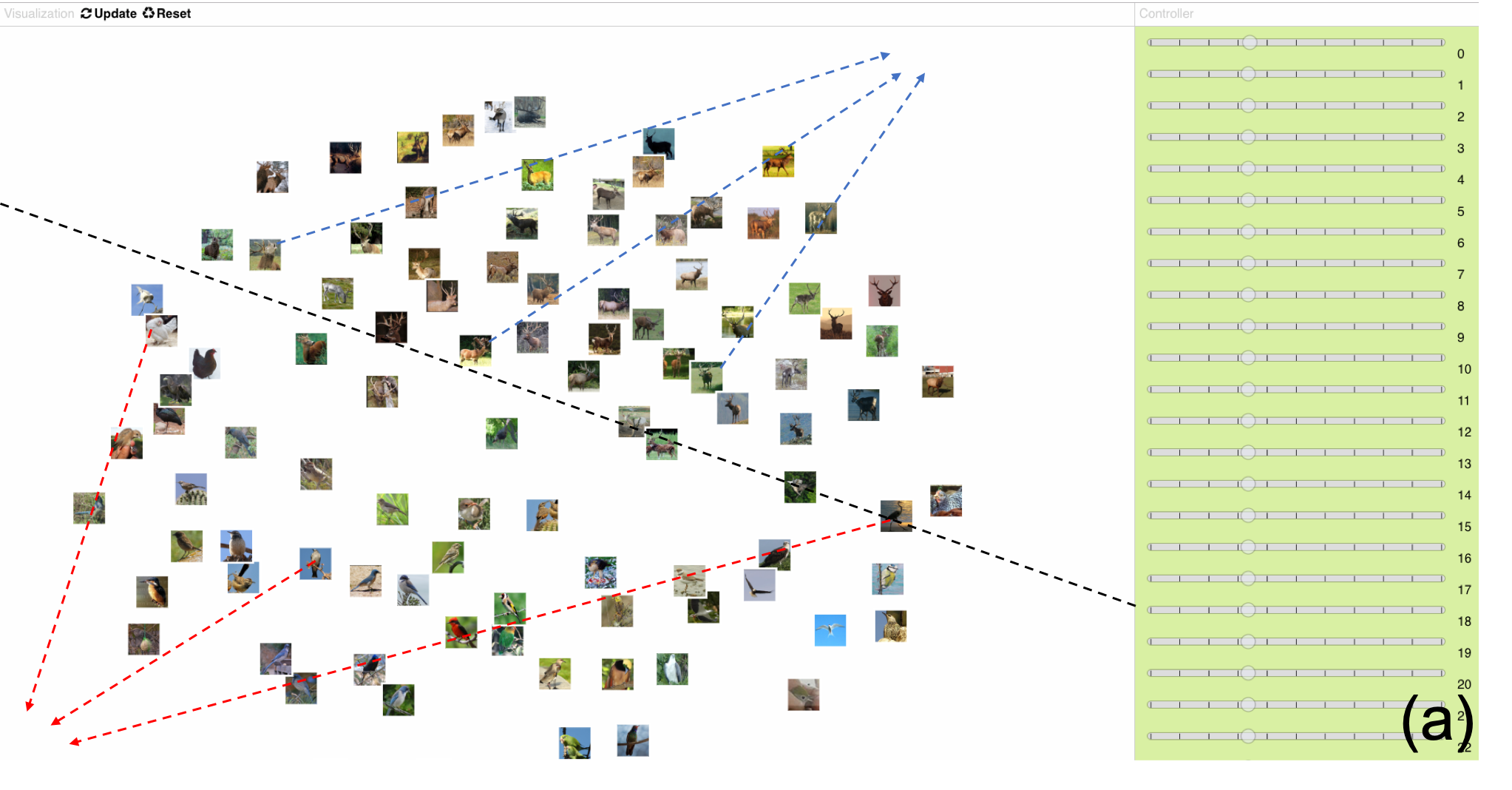}%
    \label{fig:cnn_1}
}\par\medskip
{
    \includegraphics[width=0.9\columnwidth]{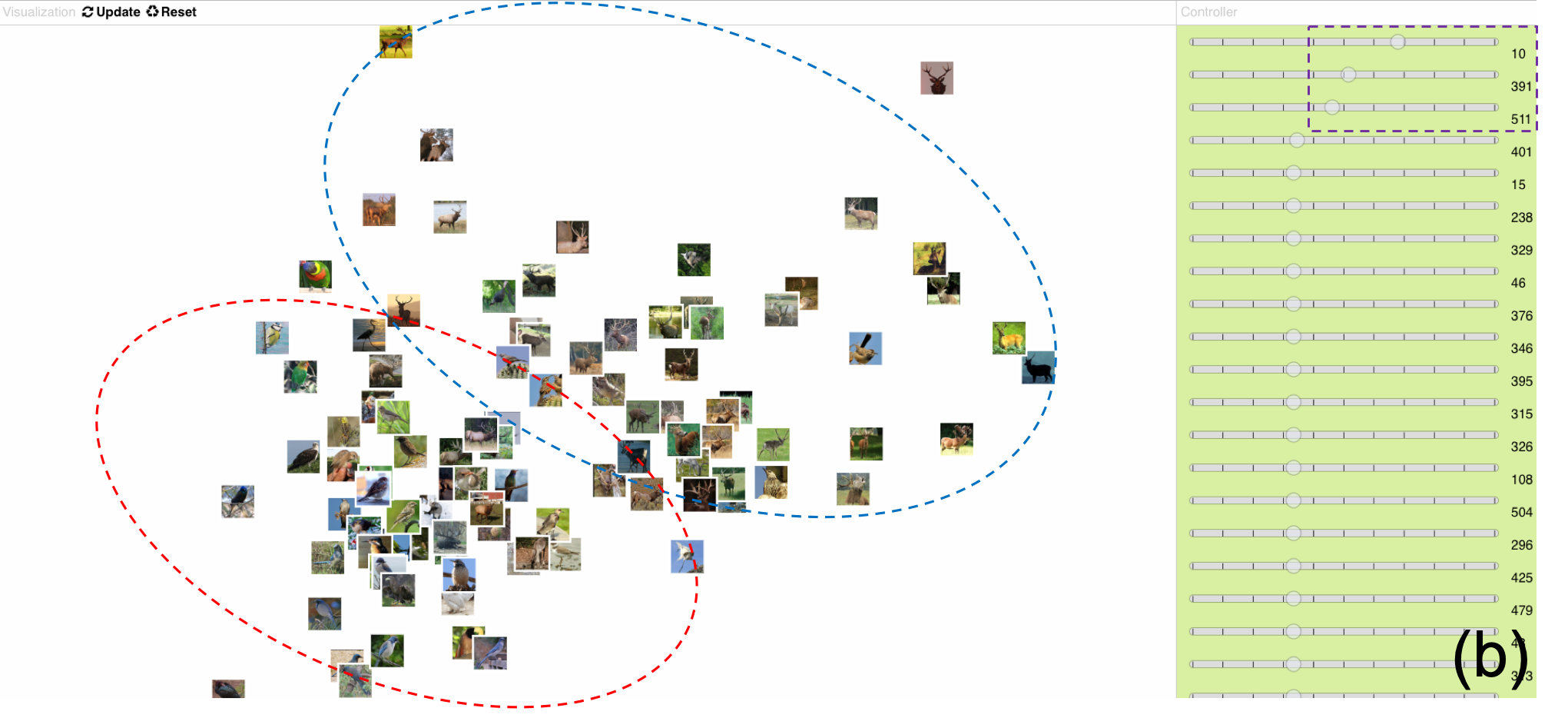}%
    \label{fig:cnn_2}
}\par\medskip        
{
    \includegraphics[width=0.9\columnwidth]{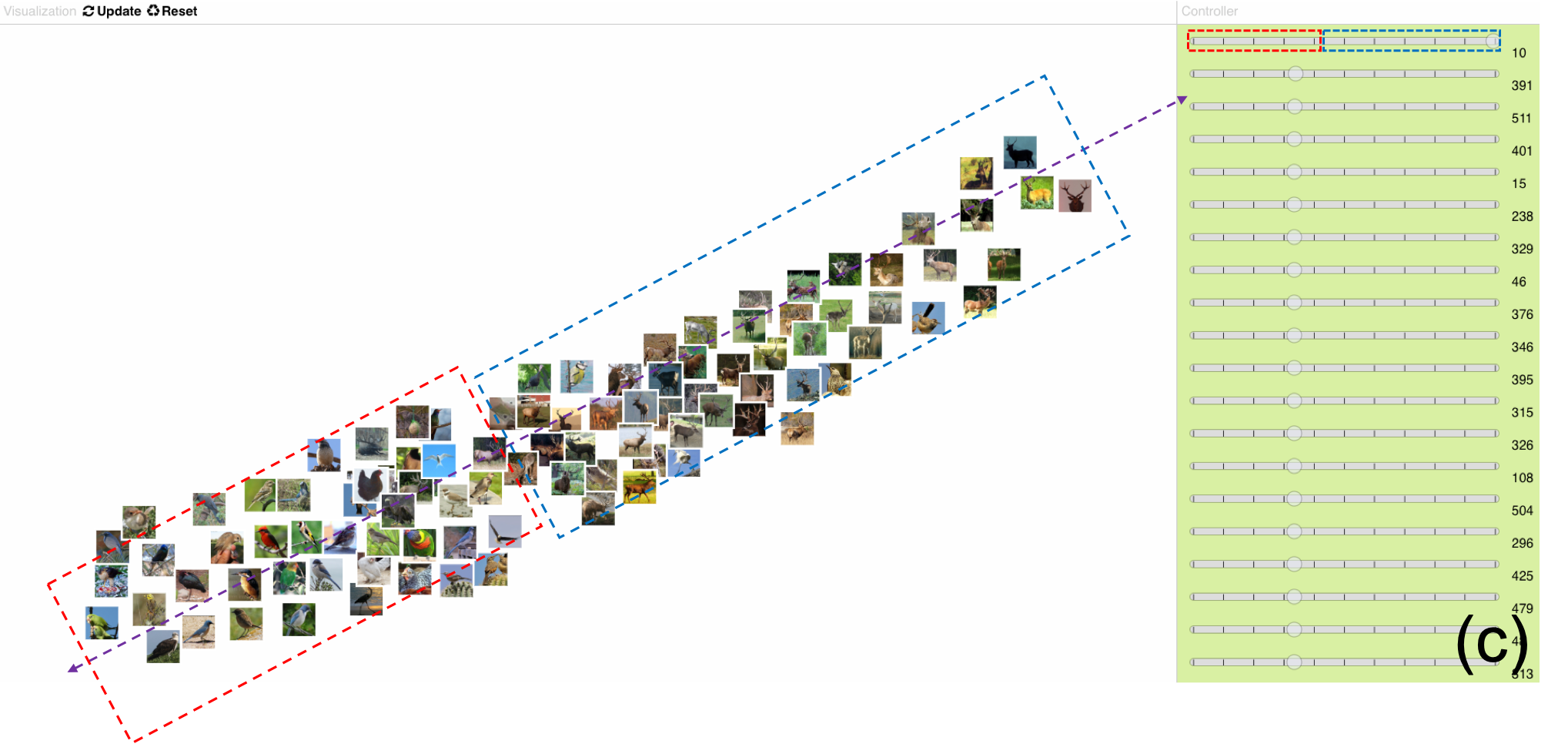}%
    \label{fig:cnn_3}
}
\caption{Several screenshots during the synthesis of the visual concept ``Deer'' using $SI_{high}$.}
\label{fig:cnn}
\end{figure}

\textit{\textbf{Initial Layout: }}
In the initial default projection~\ref{fig:cnn}a, we can identify that there is already a clear boundary between deer and birds within the projection, even though they have similar backgrounds such as trees or green ground.
We conclude that many of the learned features, when equally weighted, work together to classify deer from birds. 

\textit{\textbf{Interactive Image Movement: }}
To better express the concept ``deer,'' we interactively group three images of deer to the top right region of the projection, indicated by the blue arrows in Figure ~\ref{fig:cnn}a, and three images of other concepts far away from the deer cluster, indicated by the red arrows. Then we click the ``Update'' button to update the image layout and representation feature weights based on the moved images.

\textit{\textbf{Updated Layout: }}
After the layout updates, we identify that there are two clusters in Fig.~\ref{fig:cnn}b.
The bottom left cluster  circled in red contains images of ``birds,'' while the right cluster  circled in blue contains images of ``deer.'' 
The Representation View shows that there are at least three features of the CNN representation ($d_{10}$, $d_{391}$, and $d_{511}$) which have increased in weight.
This indicates that the visual concept ``deer'' and these three representation features are highly related. 

\textit{\textbf{Manipulating Representation Weights: }}
To further test this hypothesis, we increased and decreased the weight on the top three dimensions: $d_{10}$, $d_{391}$, $d_{511}$.
We found that $d_{10}$ could classify deer from other images, and so we maximized the weight of $d_{10}$ to further update the %image 
layout and examine the relationship between this dimension and the concept ``deer.'' 

\textit{\textbf{Conclusion: }}
As shown in Fig.~\ref{fig:cnn}c, we found that there is a linear relationship between representation dimension $d_{10}$ and the visual concept ``deer.'' 
This shows DeepVA ($SI_{high}$) could easily and effectively capture this user concept, and relate it to CNN representations.

\subsection{\texorpdfstring{$\mathbf{TAKS_{mid}: SI_{mid}}$}~\textbf{with SIFT features}}
%There are five key steps summarized as follows when carying out the study. 
\begin{figure}[tb]
\centering
{
    \includegraphics[width=0.9\columnwidth]{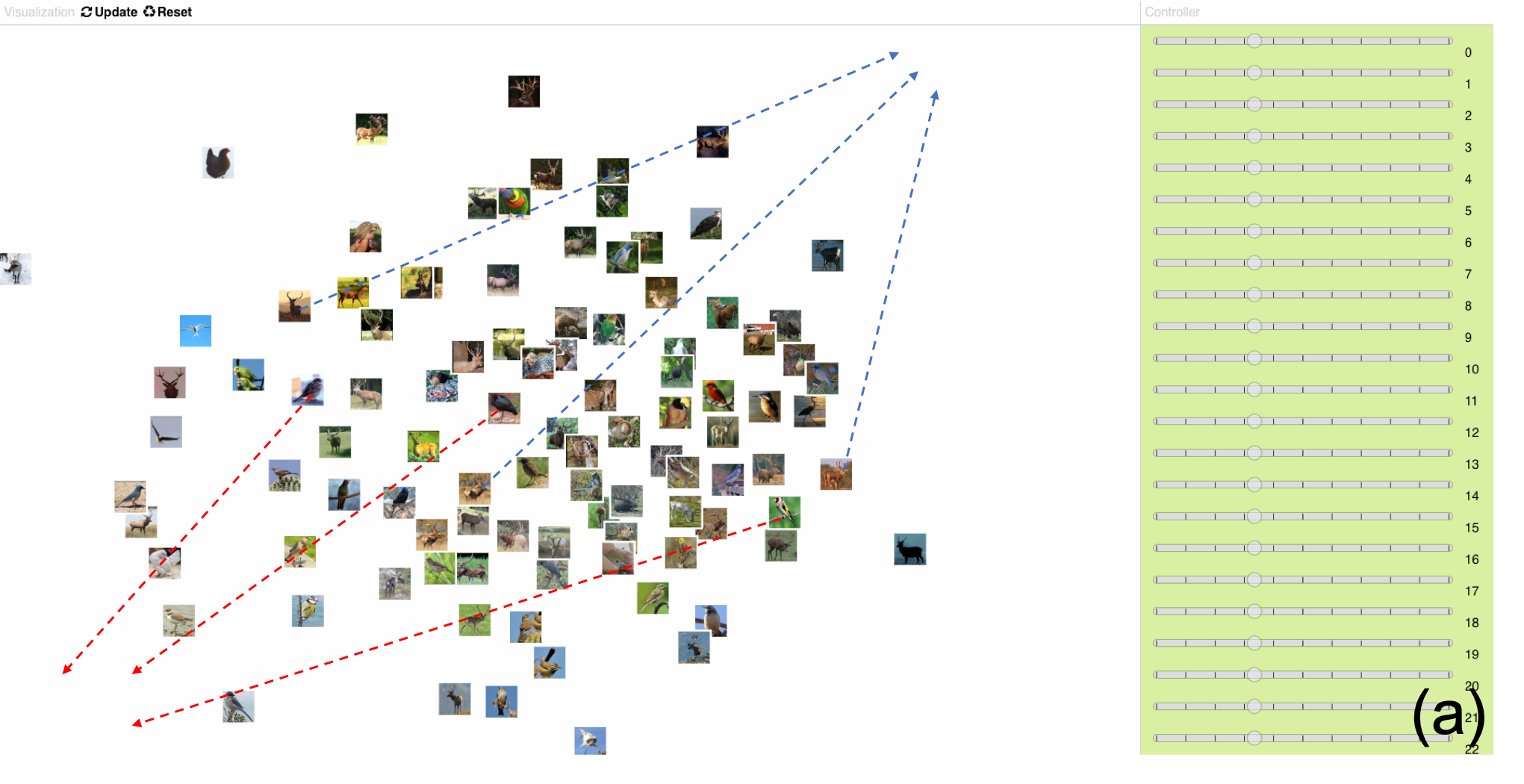}%
}\par\medskip
{
    \includegraphics[width=0.9\columnwidth]{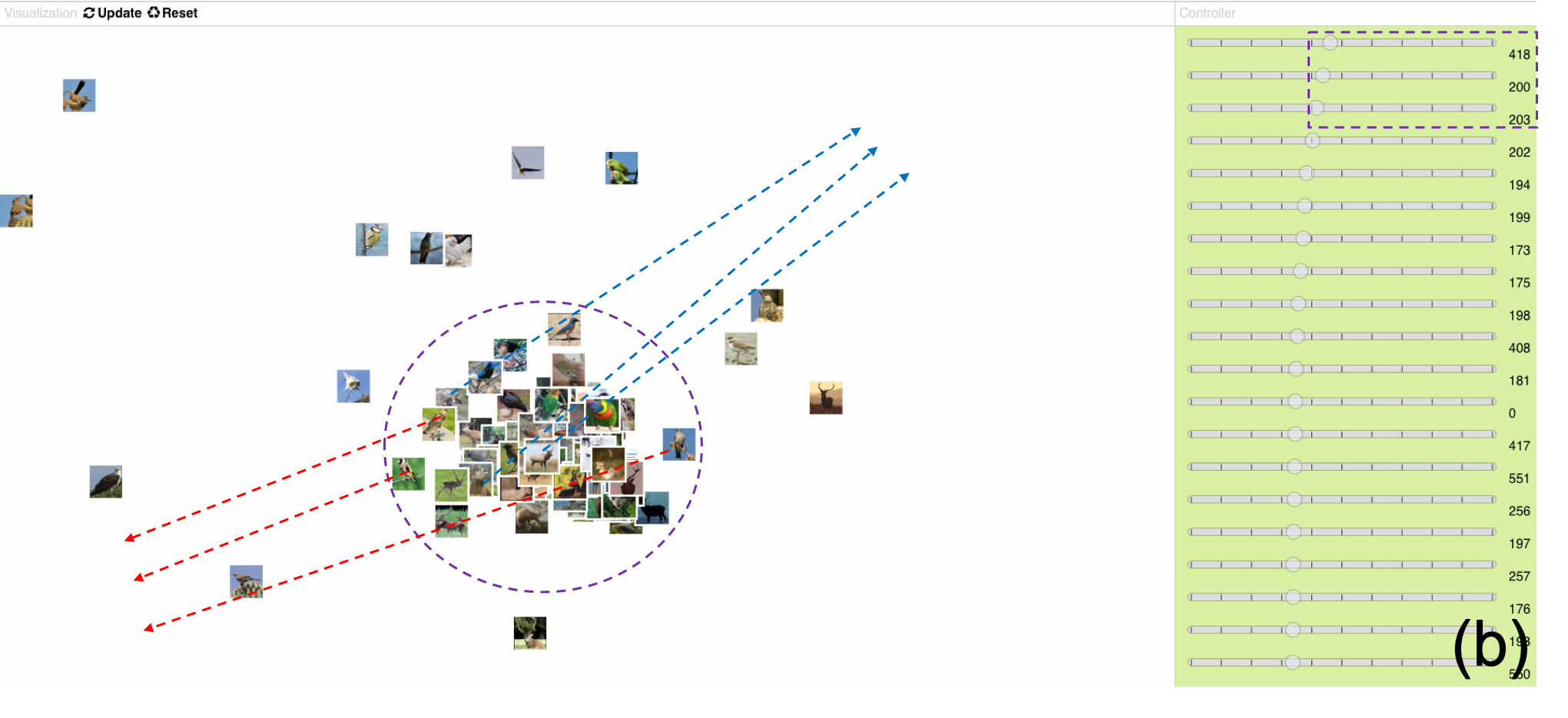}%
    \label{fig:sift_2}
}\par\medskip        
{
    \includegraphics[width=0.9\columnwidth]{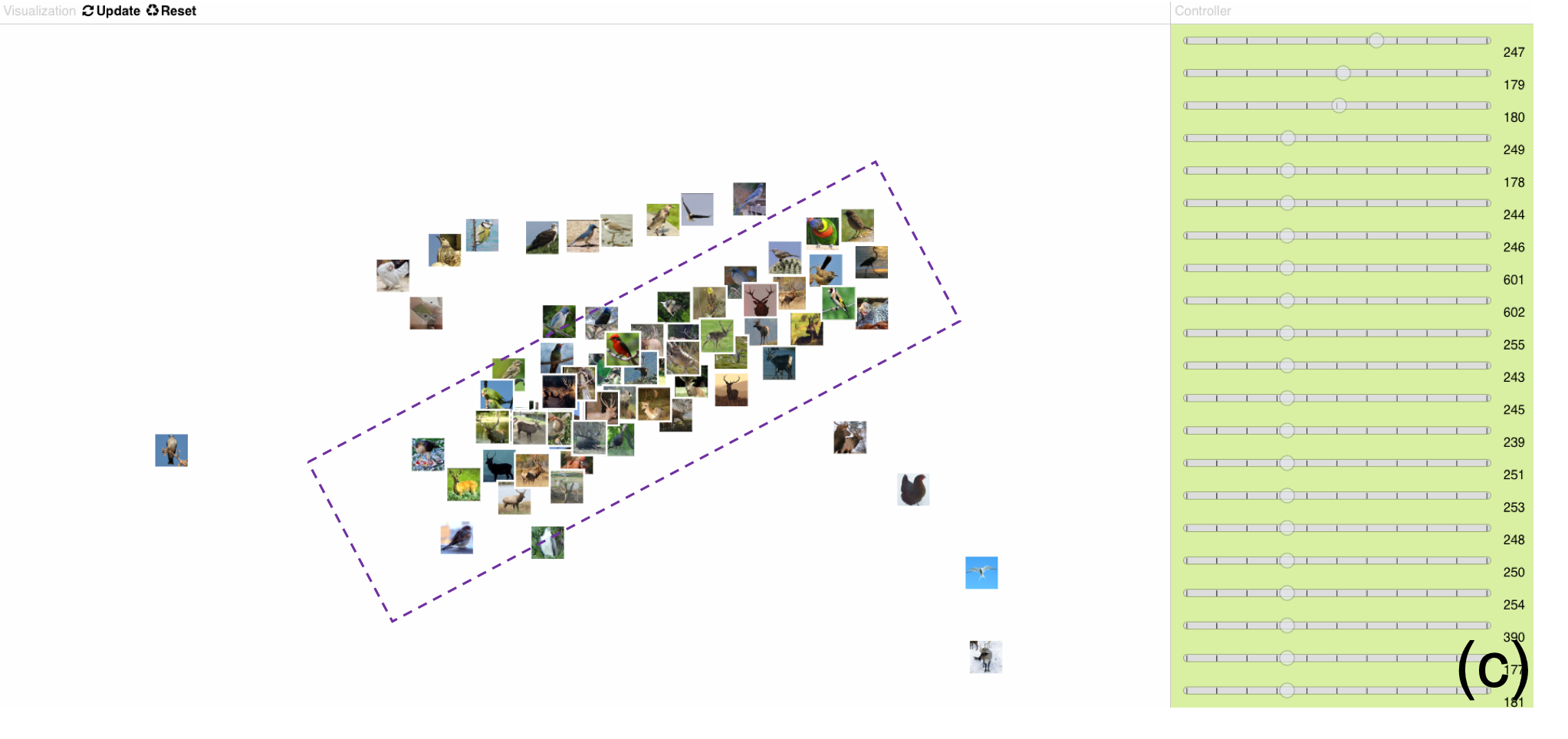}%
    \label{fig:sift_3}
}
\caption{Several screenshots during the synthesis of the visual concept ``Deer'' using $SI_{mid}$.}
\label{fig:sift}
\end{figure}

\textit{\textbf{Initial Layout:}}
Initially, the default projection~(Fig.~\ref{fig:sift}a) shows all deer and birds are scattered  in the Workspace, there is no clear boundary separating deer from others. 
It makes sense because SIFT features contain local patterns, which cannot directly distinguish two object-level concepts.

\textit{\textbf{Interactive Image Movement:}}
Similarly, we drag three images of deer to the top right region on the view, indicated by the blue arrows in Fig.~\ref{fig:sift}a, and three images without deer to the bottom left region, indicated by the red arrows, to update the model and layout.

\textit{\textbf{Updated Layout:}}
After the layout updates, we find that all deer images are grouped in the center of the view, however, mixed with some bird images~(Fig.~\ref{fig:sift}b).
We then drag six images in the cluster, to better distinguish the deer in the center cluster from others, then update layout again.

\textit{\textbf{Manipulating Feature Weights:}}
The updated layout~(\ref{fig:sift}c) shows a better cluster of deer. 
Though, there are still a small number of images about birds, especially birds on trees, mixed in the deer group. 
Then we test if the up-weighted features could be associated with the deer concept, and found that several features combined helped to group deer, but not any one particular  Sift feature.
Then more interactions are performed until a clear boundary is  shown between ``deer'' and others (shown in Fig.~\ref{fig:cost}).

\textit{\textbf{Conclusion: }}
We conclude that with SIFT features, the local features ``trees'' and ``antlers'' are similar, which leads to the mix of  ``birds in trees'' and ``deer.''
To better distinguish, more interactions and complex combinations of features are needed. 

\subsection{\texorpdfstring{$\mathbf{TASK_{mid}: SI_{low}}$}~\textbf{with Color Histogram}}

\begin{figure}[tb]
\centering
{
    \includegraphics[width=0.9\columnwidth]{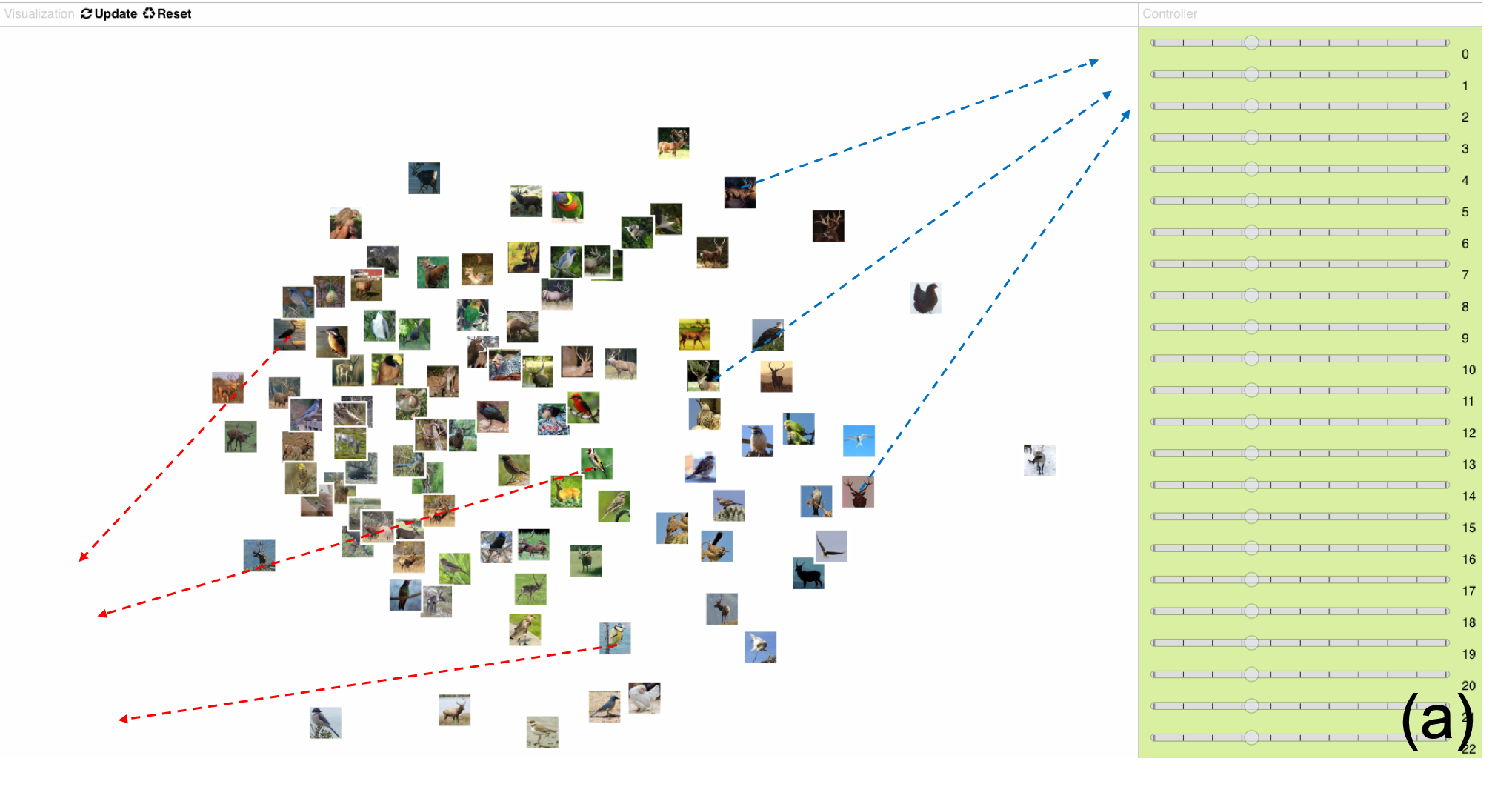}
}\par\medskip
{
    \includegraphics[width=0.9\columnwidth]{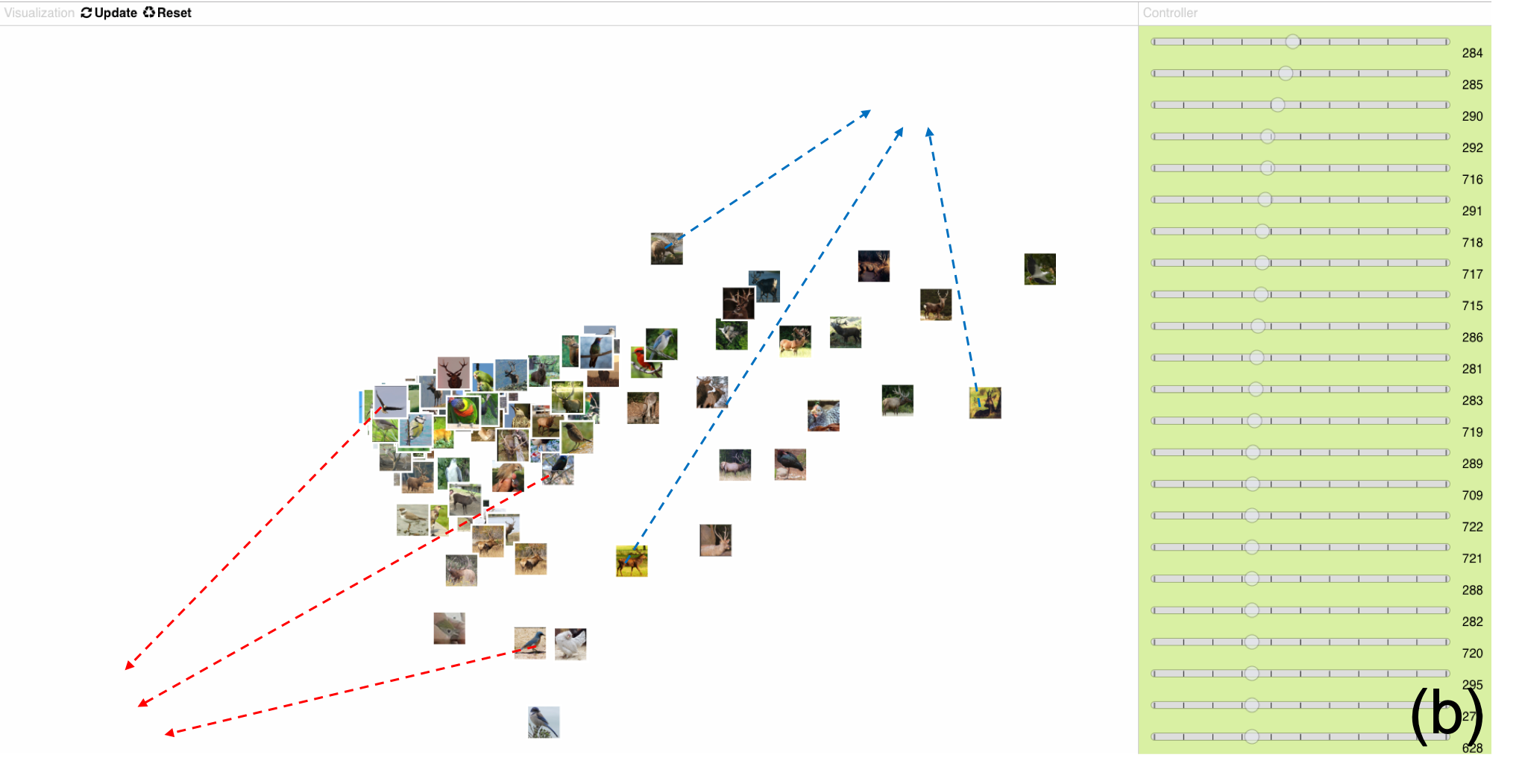}
}\par\medskip        
{
    \includegraphics[width=0.9\columnwidth]{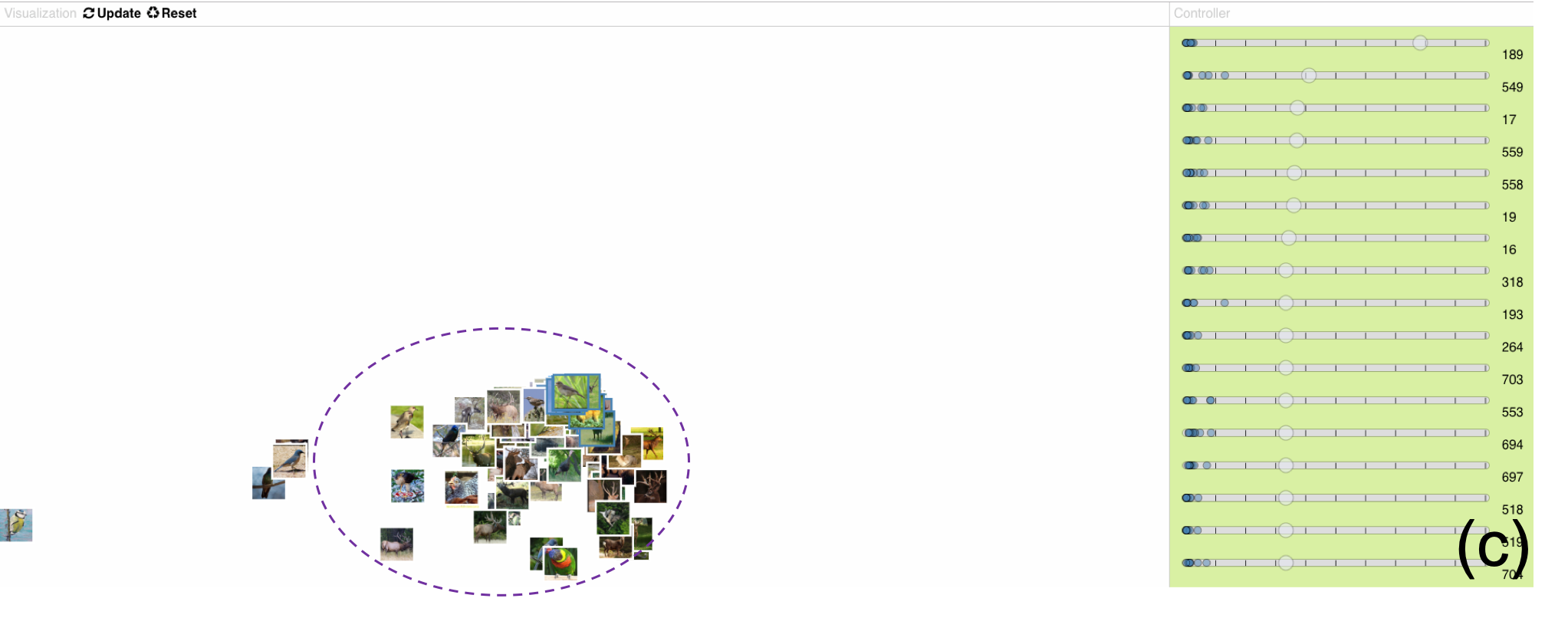}
}\par\medskip 
\caption{Several screenshots during the synthesis of the visual concept ``Deer'' using $SI_{Low}$.}
\label{fig:hist}
\end{figure}

\textit{\textbf{Initial Layout:}}
As in Fig.~\ref{fig:hist}a, images are  projected by default based on colors instead of  objects.

\textit{\textbf{Interactive Image Movement:}}
We selected and moved 6 images: three deer images to top right, indicated by the blue arrows in Fig.~\ref{fig:hist}a, and 3 other images to bottom left, indicated by the red arrows.

\textit{\textbf{Updated Layout:}}
The updated layout (Fig.~\ref{fig:hist}b) shows that images with similar colors are still closer to each other regardless of whether it contains deer.
Furthermore, all images gather together in the center, making it difficult to visually classify deer from others.
To better express the concept, we drag 6 more images and update the layout (Fig.~\ref{fig:hist}c).
The projection is worse, almost all images cluster together. 
Even with many more interactions performed, until giving up, there is still no evidence of a better structure about the concept ``deer.''

\textit{\textbf{Conclusion:}}
This evidence suggests that synthesizing complex concepts is difficult or impossible with only the basic data features in color histograms.

\section{Quantitative Results}

\subsection{Measures and Metrics}
In addition to the 3 case studies detailed above, another 6 studies were also carried out for $Task_{low}$ and $Task_{high}$. 
We evaluate the effectiveness of DeepVA~($SI_{high}$) through the 9 studies, compared with the other two methods in terms of two aspects: completeness, and interactive cost.

\subsubsection{Completeness}
As shown in Fig.~\ref{fig:tasks}, for a specified task with a specified SI method, a circle is used to represent the task completeness.
A white circle means we could not finish the task through the given method.
A gray circle means we could finish the assigned task to some degree of precision, but the  concepts in the image projection were still not completely delineated. 
A black circle means we could readily complete the task with the given SI method.

DeepVA ($SI_{high}$)  helped  complete  tasks  across all three difficulty levels.
$SI_{mid}$  mostly solved the two easier tasks ($Task_{low}$ and $Task_{mid}$) with less complex concepts, but more interactive effort was required. 
$SI_{low}$  only completed  $Task_{low}$ with a simple concept.

\subsubsection{Interactive Cost}

\begin{figure}[ht]
 \centering
 \includegraphics[width=\linewidth]{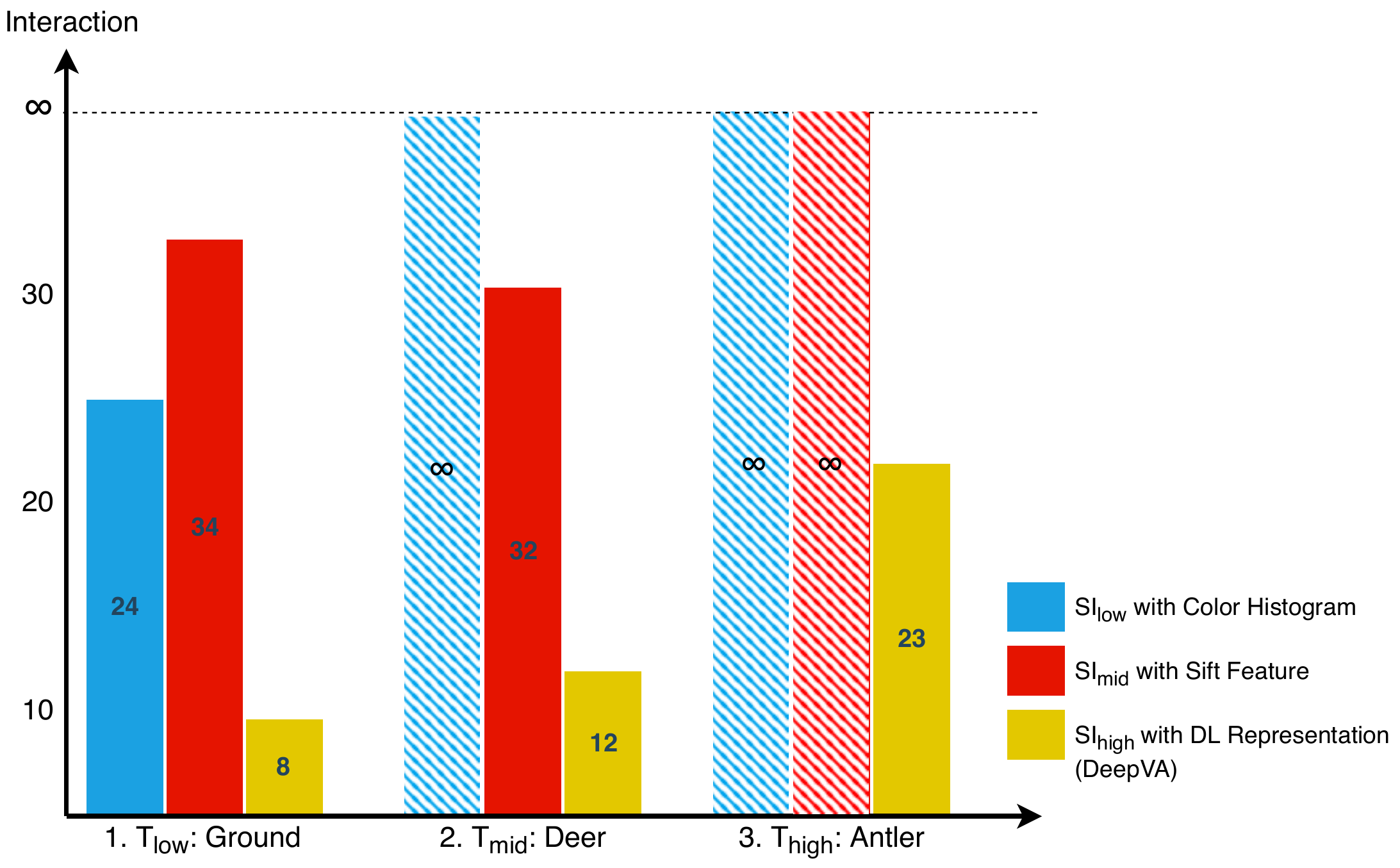}
  \caption{Interactive cost: number of interactions  required to complete the task with the assigned method.}
 \label{fig:cost}
\end{figure}

We recorded the number of user interactions (both image and feature movement) performed to complete each synthesis task as shown in Fig.~\ref{fig:cost}. 
If we gave up before successfully completing the task, interaction cost is marked as \textbf{infinity}~$\mathbf{\infty}$.
The results indicate that DeepVA  helped complete tasks with fewer  interactions, in comparison to the other two methods.
For a more complex task, more interactions were needed.
For $SI_{mid}$, in both $Task_{low}$ and $Task_{mid}$, more than 30 interactions were needed. 
It is interesting that for $Task_{low}$, $SI_{mid}$ required more interactions than $SI_{low}$.

\subsection{Summary of Results}

\subsubsection{DeepVA ($SI_{high}$)  is more effective}
\begin{figure}[h]
 \centering % avoid the use of \begin{center}...\end{center} and use \centering instead (more compact)
 \includegraphics[width=\columnwidth]{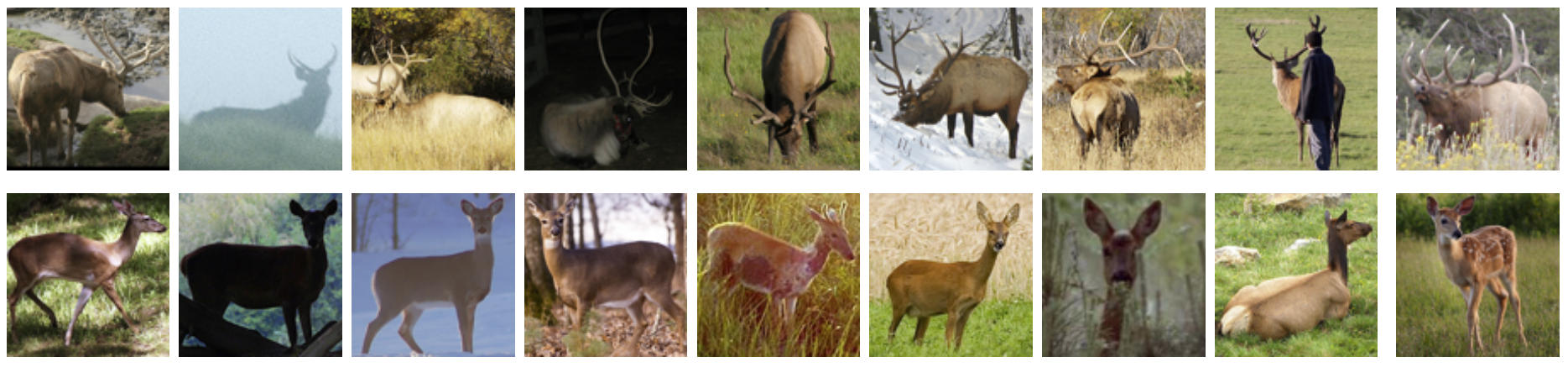}
  \caption{The most and least similar pictures of `antler' concept over $d_{244}$. The upper row of nine images are deer with the largest values of $d_{244}$, while the lower row contains the nine deer images with the smallest values of $d_{244}$.  We can conclude that large values of $d_{244}$ imply the presence of antlers.
  }
 \label{fig:sorted-antler}
\end{figure}

The result in Fig.~\ref{fig:tasks}~shows an upper triangular structure that well-matches our hypothesis regarding the match-up of task and feature abstraction levels. DeepVA is able to effectively complete all three tasks.
This indicates that higher-level features can enable users to efficiently synthesize  higher-level  concepts (as well as simpler low-level concepts) using OLI. 
Even for $Task_{high}$, with the most complex concept ``antler,'' DeepVA managed to map it to one representation feature ($d_{244}$ as shown in Fig.~\ref{fig:vis} and Fig.~\ref{fig:sorted-antler}). 
This can greatly reduce users' interaction effort during sensemaking, and narrow the gap  between complex cognitive concepts and  high-dimensional computational features.
In contrast to DeepVA, $SI_{mid}$ and $SI_{low}$ cannot adequately capture the complex concepts expressed in $Task_{high}$ with lower-level features. 
Even in $Task_{low}$ and $Task_{mid}$, several features together were needed to define the visual concept and more interactive training was required.
Analysts may need to carefully tweak the combination of features to work for the desired concept.
This clearly answered our research question that DeepVA~($SI_{high}$) can improve SI performance at synthesizing concepts at all levels. 

\subsubsection{DeepVA ($SI_{high}$) is more efficient}
As shown in Fig.~\ref{fig:cost}, in each case study,  DeepVA required fewer interactions  than other methods.
OLI with higher-level features, is more efficient in capturing users' synthesized concepts. 
The number of interactions used in $Task_{high}$ with DeepVA is  less than the other two methods used in  simpler $Task_{low}$ and $Task_{mid}$. 
Furthermore, the relationship between interactive cost and the task complexity in DeepVA is more consistent than the other methods. That is, the more complex a task is, the more interactive cost is needed. 
The other two methods seem to be more tuned to a specific task. 
$SI_{low}$ performs best on $Task_{low}$, while $SI_{mid}$ performs best on $Task_{mid}$.

\section{Discussion}
Through these studies, we find that OLI with higher-level features can capture more complex concepts  more efficiently.
OLI with deep learning representations (DeepVA) can effectively and efficiently help users to synthesize complex concepts. % through SI with deep learning representations.
This also means transfer learning does extract meaningful and appropriate representations from the given image dataset, using the pre-trained deep learning model.
In addition to these research results, we also find the following insights. 

\subsection{Learned Features as Representative of Cognition}
Through these case studies, we found that all three visual concepts of different complexity can be directly mapped to one feature of DL representations. 
In $Task_{low}$, the concept ``ground'' can be defined by $d_{184}$.
In $Task_{mid}$, the concept ``deer '' is directly mapped to $d_{10}$~(Fig.~\ref{fig:cnn}c).
In $Task_{high}$, the concept ``antler'' is linked to $d_{244}$~(Fig.~\ref{fig:vis}). 
Cognitive intents could be directly mapped to deep learning features. 
This suggests a potential match-up between cognitive and computational DL features.

It was somewhat surprising that DeepVA effectively captured cognitive concepts at all three levels of complexity, not just the high level.
The match-up not only explains the effectiveness of DeepVA, but also illustrates the internal learning process of DL to discover data representations at multiple levels of abstraction.
With a huge amount of data labelled for  specific tasks such as feature detection or classification, deep learning techniques can automatically discover the best representations for  the tasks. 
Through this process, DL compresses and distributes data from input layers into later layers. 
In the ImageNet dataset, many kinds of images are labelled in many kinds of categories as training data, spanning many kinds of cognitive tasks at many different levels.
Thus, the learned DL captures all relevant concepts across different complexity levels, such as ``ground'',    ``deer'', and ``antlers''.

\subsection{Coupling cognition and computation with DeepVA}
\begin{figure}[tb]
\centering
  \includegraphics[width=1.0\columnwidth]{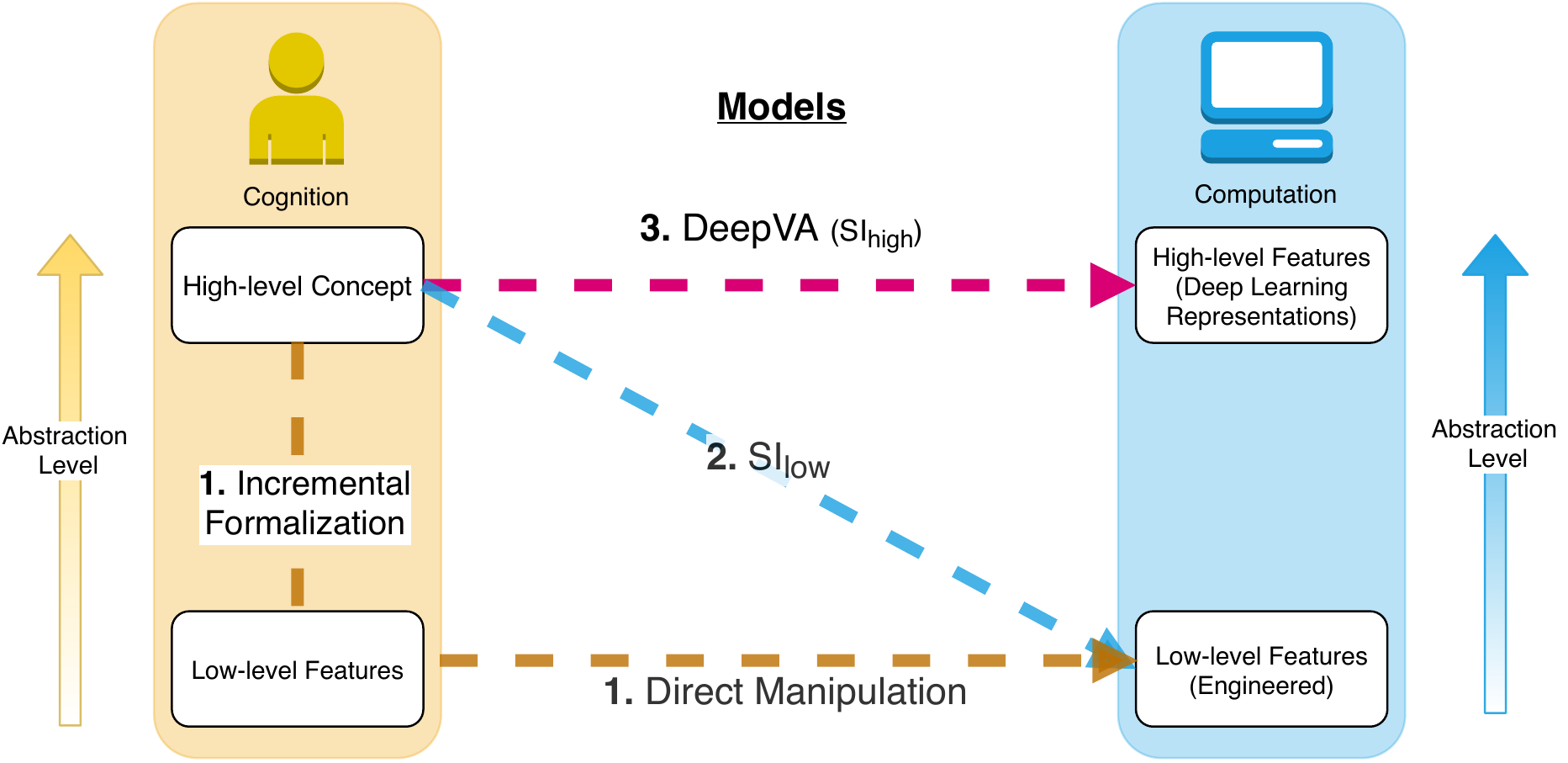}
  \caption{Coupling cognition and computation through SI methods: 
  (Method 1) Users internally map high-level concepts to low-level features, then directly manipulate those engineered features. 
  (Method 2) $SI_{low}$ maps user's high-level concepts to engineered low-level features, via various SI methods.
  (Method 3) $SI_{high}$~(DeepVA) maps user's high-level concepts directly to high-level learned features, via various SI methods. 
}
  ~\label{fig:gap}
\end{figure}
%\chris{Fig3: i dont understand linear vs non-linear. is that important distinction?  also i think #1 should be Direct Manip.}

Deep learning techniques could greatly enhance SI systems. 
We  analyze  the advantage of SIs with deep learning from the perspective of bridging the gap between cognition and computation for complex tasks.
As shown in Fig.~\ref{fig:gap}, users' cognitive reasoning can be represented through a range of abstraction levels. Through the process of incremental formalization, users can gradually internally transform their higher-level abstract concepts into specific lower-level features such as keywords or image colors.  A similar abstraction hierarchy exists on the computational side, ranging from traditional low-level engineered features, like pixels, to higher-level abstractions like deep learning representations. A goal of VA is to bridge the chasm between cognition and computation via various models.

Traditionally, \textbf{direct manipulation} forces users to first formalize their concepts into low-level features, so that they can then directly manipulate the corresponding engineered features. For example, Google requires users to first identify specific keywords to enter, and IN-SPIRE~\cite{1382935} enables users to change the weighting of specific keywords when computing the corpus projection.
This overhead is especially problematic when users are already devoting their full cognitive effort to sensemaking tasks. At early stages in the sensemaking process, high-level cognitive concepts can be difficult for users to express and formalize in terms of unfamiliar low-level features.

As discussed in Sec.~\ref{sec-VA}, \textbf{semantic interactions such as observation-level interaction} help the process by
extrapolating from the user's synthesis-related interactions to map the user's high-level cognitive reasoning concepts into the low-level computational feature space. An advantage of this strategy is that it can help users perform incremental formalism~\cite{shipman1994supporting} by automatically mapping to the low-level features. Endert et al.~\cite{Endert:he} showed the distinction between user's high-level concepts and computational low-level features in their user study.
However, SI is thus inherently limited by the chosen feature set. 
To date, SI with low-level engineered feature sets (method 2 in Fig.~\ref{fig:gap}),  have been used to perform mappings from simple linear models (e.g. ForceSPIRE~\cite{endert2012semantic}) to more complex non-linear models (e.g. AxiSketcher~\cite{7534876}, iGTM~\cite{10.1371/journal.pone.0129122}).  
However, regardless of the complexity of the mapping method, the use of low-level feature sets will always limit the inference capability.
Like other machine learning techniques, without enough knowledge pre-loaded into the features, it is difficult to infer users' intents based on visual interactions. 

Different from the above two methods, \textbf{DeepVA} involves identifying and overcoming this limitation.
DeepVA improves the cognitive-computational coupling by using deep learning representations as high-level features, which update the model with higher-level conceptual semantics (method 3 in Fig.~\ref{fig:gap}).
Unlike lower-level engineered features, deep learning representations could contain more high-level concepts. 
Thus, even basic linear mappings between DL features and human cognition is valid and enough for SI models.

\subsection{Interpretable Deep Learning Representations}
We mainly explored how VA powered by deep learning can facilitate analysts in complex synthesis tasks.
Beyond its usage in VA, this combination can also assist DL researchers in interpreting DL representations by adding human-centered tasks as constraints. DeepVA can help understand the semantic meanings behind the otherwise mysterious DL features.
As in the case study,  mappings users' interactions onto DL features, and the use of the features to help users explore more about specific semantic concepts, can both help DL researchers.
Different image projections can express concepts, and reveal insights about the meaning of the associated up-weighted DL features.
Likewise, through DeepVA, users can make sense of a single feature by up-weighting it, and then examining the resulting projection. 
Semantic meanings of the up-weighted feature could be expressed and understood through the updated Workspace view. 
For example, to understand the feature $d_{244}$, we maximized its weight by dragging its slider to the right.
Through the updated image layout in Fig.~\ref{fig:vis}, we could conclude that the feature $d_{244}$ is related to the visual concept ``antler.''

\subsection{Limitations and Future Work}

%However,
In its current state, DeepVA might not help 
users with incremental formalism~\cite{shipman1994supporting}, because both the cognitive and computational activity are occurring at higher-levels of abstraction. An open question is to investigate how the high-level DL features can be automatically mapped back to low-level features to assist the user in completing the formalization loop.

DeepVA is highly dependent on pre-trained DL networks through transfer learning techniques.
If the dataset for analysis is too different from the big dataset that was used to train the deep learning models, the extracted representations might not contain meaningful concepts that users are interested in.
Additional research is needed to enable interactive re-training of the deep learning models that would shift it towards users' concepts of interest.
Furthermore, significantly greater scalability is needed.
To provide a VA application, supported fully by deep learning models, would require thousands of features for representations.
This problem can potentially be alleviated by  selecting the most relevant features and layers of DL representations based on the targeted analysis tasks during the transfer learning process.
An interesting question for future work is how the selection of DL layers could adapt to human incremental formalization of concepts. For example, lower layers might be selected as the user's concepts become more formalized.

\section{Conclusion}
In this paper, we explored  SI powered by high-level DL representations, in comparison to traditional lower-level engineered data features. 
The revised SI model, called DeepVA, consists of two methods: (1) transfer learning to integrate deep learning into VA, (2) observation-level interaction with  high-level DL features to capture and model users' cognitive reasoning process.
To evaluate DeepVA, specifically how SI with higher-level data features overcomes the limitations of SI with low-level features, we implemented an SI system for visual image analysis that enables multiple data feature sets.
We then performed 9 case studies on interactive synthesis tasks at three different conceptual abstraction levels, using SI methods with three different feature abstraction levels. 
Overall, DL features improved SI performance.
More generally, higher-level  data features  better support SI users in constructing higher-level concepts.
Results demonstrated that SI with higher-level DL features can better capture users' complex cognitive syntheses with less interactive cost than traditional lower-level data features.
As a result, DeepVA is more effective and efficient in supporting interactive sensemaking tasks via SI.
Furthermore, DeepVA offers new insight into effectively bridging the gap between human cognition and computational models.

%% if specified like this the section will be committed in review mode
% \acknowledgments{
% The authors wish to thank A, B, C. This work was supported in part by
% a grant from XYZ.}

%\bibliographystyle{abbrv}
\bibliographystyle{abbrv-doi}

\bibliography{template}

\begin{thebibliography}{10}

\bibitem{abadi2016tensorflow}
M.~Abadi, A.~Agarwal, P.~Barham, E.~Brevdo, Z.~Chen, C.~Citro, G.~S. Corrado,
  A.~Davis, J.~Dean, M.~Devin, et~al.
\newblock Tensorflow: Large-scale machine learning on heterogeneous distributed
  systems.
\newblock {\em arXiv preprint arXiv:1603.04467}, 2016.

\bibitem{athiwaratkun2015feature}
B.~Athiwaratkun and K.~Kang.
\newblock Feature representation in convolutional neural networks.
\newblock {\em arXiv preprint arXiv:1507.02313}, 2015.

\bibitem{bergstra2010theano}
J.~Bergstra, O.~Breuleux, F.~Bastien, P.~Lamblin, R.~Pascanu, G.~Desjardins,
  J.~Turian, D.~Warde-Farley, and Y.~Bengio.
\newblock Theano: A cpu and gpu math compiler in python.
\newblock In {\em Proc. 9th Python in Science Conf}, vol.~1, 2010.

\bibitem{Bradel:dd}
L.~Bradel, C.~North, L.~House, and S.~Leman.
\newblock {Multi-model semantic interaction for text analytics}.
\newblock In {\em 2014 IEEE Conference on Visual Analytics Science and
  Technology (VAST)}, pp. 163--172. IEEE.

\bibitem{8265023}
H.~{Cheng}, A.~{Cardone}, S.~{Jain}, E.~{Krokos}, K.~{Narayan},
  S.~{Subramaniam}, and A.~{Varshney}.
\newblock Deep-learning-assisted volume visualization.
\newblock {\em IEEE Transactions on Visualization and Computer Graphics},
  25(2):1378--1391, Feb 2019. doi: {{%
10\hspace{.1pt}\discretionary{.}{%
}{.}\hspace{.4pt}1109\discretionary{/}{%
}{/}TVCG\hspace{.1pt}\discretionary{.}{%
}{.}\hspace{.4pt}2018\hspace{.1pt}\discretionary{.}{%
}{.}\hspace{.4pt}2796085}}


\bibitem{8402187}
J.~Choo and S.~Liu.
\newblock Visual analytics for explainable deep learning.
\newblock {\em IEEE Computer Graphics and Applications}, 38(04):84--92, jul
  2018. doi: {{%
10\hspace{.1pt}\discretionary{.}{%
}{.}\hspace{.4pt}1109\discretionary{/}{%
}{/}MCG\hspace{.1pt}\discretionary{.}{%
}{.}\hspace{.4pt}2018\hspace{.1pt}\discretionary{.}{%
}{.}\hspace{.4pt}042731661}}


\bibitem{coates2011analysis}
A.~Coates, A.~Ng, and H.~Lee.
\newblock An analysis of single-layer networks in unsupervised feature
  learning.
\newblock In {\em Proceedings of the fourteenth international conference on
  artificial intelligence and statistics}, pp. 215--223, 2011.

\bibitem{7160906}
A.~Endert, R.~Chang, C.~North, and M.~Zhou.
\newblock Semantic interaction: Coupling cognition and computation through
  usable interactive analytics.
\newblock {\em IEEE Computer Graphics and Applications}, 35(4):94--99, July
  2015. doi: {{%
10\hspace{.1pt}\discretionary{.}{%
}{.}\hspace{.4pt}1109\discretionary{/}{%
}{/}MCG\hspace{.1pt}\discretionary{.}{%
}{.}\hspace{.4pt}2015\hspace{.1pt}\discretionary{.}{%
}{.}\hspace{.4pt}91}}


\bibitem{Endert:he}
A.~Endert, P.~Fiaux, and C.~North.
\newblock Semantic interaction for sensemaking: inferring analytical reasoning
  for model steering.
\newblock {\em IEEE Transactions on Visualization and Computer Graphics},
  18(12):2879--2888, 2012.

\bibitem{endert2012semantic}
A.~Endert, P.~Fiaux, and C.~North.
\newblock Semantic interaction for visual text analytics.
\newblock In {\em Proceedings of the SIGCHI conference on Human factors in
  computing systems}, pp. 473--482. ACM, 2012.

\bibitem{Endert:ji}
A.~Endert, C.~Han, D.~Maiti, L.~House, S.~Leman, and C.~North.
\newblock {Observation-level interaction with statistical models for visual
  analytics}.
\newblock In {\em 2011 IEEE Conference on Visual Analytics Science and
  Technology (VAST)}, pp. 121--130. IEEE.

\bibitem{Girshick:vu}
R.~Girshick, J.~Donahue, T.~Darrell, and J.~Malik.
\newblock Rich feature hierarchies for accurate object detection and semantic
  segmentation.
\newblock In {\em Proceedings of the IEEE conference on computer vision and
  pattern recognition}, pp. 580--587, 2014.

\bibitem{391417}
J.~Hafner, H.~S. Sawhney, W.~Equitz, M.~Flickner, and W.~Niblack.
\newblock Efficient color histogram indexing for quadratic form distance
  functions.
\newblock {\em IEEE Transactions on Pattern Analysis and Machine Intelligence},
  17(7):729--736, July 1995. doi: {{%
10\hspace{.1pt}\discretionary{.}{%
}{.}\hspace{.4pt}1109\discretionary{/}{%
}{/}34\hspace{.1pt}\discretionary{.}{%
}{.}\hspace{.4pt}391417}}


\bibitem{10.1371/journal.pone.0129122}
C.~Han, L.~House, and S.~C. Leman.
\newblock Expert-guided generative topographical modeling with visual to
  parametric interaction.
\newblock {\em PLOS ONE}, 11(2):1--14, 02 2016. doi: {{%
10\hspace{.1pt}\discretionary{.}{%
}{.}\hspace{.4pt}1371\discretionary{/}{%
}{/}journal\hspace{.1pt}\discretionary{.}{%
}{.}\hspace{.4pt}pone\hspace{.1pt}\discretionary{.}{%
}{.}\hspace{.4pt}0129122}}


\bibitem{he2016deep}
K.~He, X.~Zhang, S.~Ren, and J.~Sun.
\newblock Deep residual learning for image recognition.
\newblock In {\em Proceedings of the IEEE conference on computer vision and
  pattern recognition}, pp. 770--778, 2016.

\bibitem{DBLP:journals/corr/HodasE16}
N.~O. Hodas and A.~Endert.
\newblock Adding semantic information into data models by learning domain
  expertise from user interaction.
\newblock {\em CoRR}, abs/1604.02935, 2016.

\bibitem{hohman2018visual}
F.~Hohman, M.~Kahng, R.~Pienta, and D.~H. Chau.
\newblock Visual analytics in deep learning: An interrogative survey for the
  next frontiers.
\newblock {\em IEEE Transactions on Visualization and Computer Graphics}, 2018.

\bibitem{house2015bayesian}
L.~House, S.~Leman, and C.~Han.
\newblock Bayesian visual analytics: Bava.
\newblock {\em Stat. Anal. Data Min.}, 8(1):1--13, Feb. 2015. doi: {{%
10\hspace{.1pt}\discretionary{.}{%
}{.}\hspace{.4pt}1002\discretionary{/}{%
}{/}sam\hspace{.1pt}\discretionary{.}{%
}{.}\hspace{.4pt}11253}}


\bibitem{itseez2015opencv}
Itseez.
\newblock Open source computer vision library.
\newblock \url{https://github.com/itseez/opencv}, 2015.

\bibitem{cook2005illuminating}
J.~J~Thomas, K.~A~Cook, I.~Electrical, and E.~Engineers.
\newblock Illuminating the path: The research and development agenda for visual
  analytics.
\newblock Technical report, 01 2005.

\bibitem{jia2014caffe}
Y.~Jia, E.~Shelhamer, J.~Donahue, S.~Karayev, J.~Long, R.~Girshick,
  S.~Guadarrama, and T.~Darrell.
\newblock Caffe: Convolutional architecture for fast feature embedding.
\newblock In {\em Proceedings of the 22nd ACM international conference on
  Multimedia}, pp. 675--678. ACM, 2014.

\bibitem{kim2017interpretability}
B.~Kim, M.~Wattenberg, J.~Gilmer, C.~Cai, J.~Wexler, F.~Viegas, and R.~Sayres.
\newblock Interpretability beyond feature attribution: Quantitative testing
  with concept activation vectors (tcav).
\newblock {\em arXiv preprint arXiv:1711.11279}, 2017.

\bibitem{krizhevsky2012imagenet}
A.~Krizhevsky, I.~Sutskever, and G.~E. Hinton.
\newblock Imagenet classification with deep convolutional neural networks.
\newblock In {\em Advances in neural information processing systems}, pp.
  1097--1105, 2012.

\bibitem{Krizhevsky:2012:ICD:2999134.2999257}
A.~Krizhevsky, I.~Sutskever, and G.~E. Hinton.
\newblock Imagenet classification with deep convolutional neural networks.
\newblock In {\em Proceedings of the 25th International Conference on Neural
  Information Processing Systems - Volume 1}, NIPS'12, pp. 1097--1105. Curran
  Associates Inc., USA, 2012.

\bibitem{7534876}
B.~C. {Kwon}, H.~{Kim}, E.~{Wall}, J.~{Choo}, H.~{Park}, and A.~{Endert}.
\newblock Axisketcher: Interactive nonlinear axis mapping of visualizations
  through user drawings.
\newblock {\em IEEE Transactions on Visualization and Computer Graphics},
  23(1):221--230, Jan 2017. doi: {{%
10\hspace{.1pt}\discretionary{.}{%
}{.}\hspace{.4pt}1109\discretionary{/}{%
}{/}TVCG\hspace{.1pt}\discretionary{.}{%
}{.}\hspace{.4pt}2016\hspace{.1pt}\discretionary{.}{%
}{.}\hspace{.4pt}2598446}}


\bibitem{LeCun:2015dt}
Y.~LeCun, Y.~Bengio, and G.~Hinton.
\newblock Deep learning.
\newblock {\em nature}, 521(7553):436, 2015.

\bibitem{Leman:2013it}
S.~C. Leman, L.~House, D.~Maiti, A.~Endert, and C.~North.
\newblock {Visual to Parametric Interaction (V2PI)}.
\newblock {\em PLOS ONE}, 8(3):e50474, Mar. 2013.

\bibitem{7536654}
M.~Liu, J.~Shi, Z.~Li, C.~Li, J.~Zhu, and S.~Liu.
\newblock Towards better analysis of deep convolutional neural networks.
\newblock {\em IEEE Transactions on Visualization and Computer Graphics},
  23(1):91--100, Jan 2017. doi: {{%
10\hspace{.1pt}\discretionary{.}{%
}{.}\hspace{.4pt}1109\discretionary{/}{%
}{/}TVCG\hspace{.1pt}\discretionary{.}{%
}{.}\hspace{.4pt}2016\hspace{.1pt}\discretionary{.}{%
}{.}\hspace{.4pt}2598831}}


\bibitem{Lowe2004}
D.~G. Lowe.
\newblock Distinctive image features from scale-invariant keypoints.
\newblock {\em International Journal of Computer Vision}, 60(2):91--110, Nov
  2004. doi: {{%
10\hspace{.1pt}\discretionary{.}{%
}{.}\hspace{.4pt}1023\discretionary{/}{%
}{/}B\discretionary{:}{%
}{:}VISI\hspace{.1pt}\discretionary{.}{%
}{.}\hspace{.4pt}0000029664\hspace{.1pt}\discretionary{.}{%
}{.}\hspace{.4pt}99615\hspace{.1pt}\discretionary{.}{%
}{.}\hspace{.4pt}94}}


\bibitem{paszke2017automatic}
A.~Paszke, S.~Gross, S.~Chintala, G.~Chanan, E.~Yang, Z.~DeVito, Z.~Lin,
  A.~Desmaison, L.~Antiga, and A.~Lerer.
\newblock Automatic differentiation in pytorch.
\newblock 2017.

\bibitem{pirolli_2005}
P.~Pirolli and S.~Card.
\newblock The sensemaking process and leverage points for analyst technology as
  identified through cognitive task analysis.
\newblock pp. 2--4, 2005.

\bibitem{DBLP:journals/corr/abs-1802-05316}
M.~Pirrung, N.~Hilliard, A.~Yankov, N.~O'Brien, P.~Weidert, C.~D. Corley, and
  N.~O. Hodas.
\newblock Sharkzor: Interactive deep learning for image triage, sort and
  summary.
\newblock {\em CoRR}, abs/1802.05316, 2018.

\bibitem{razavian2014cnn}
A.~S. Razavian, H.~Azizpour, J.~Sullivan, and S.~Carlsson.
\newblock Cnn features off-the-shelf: an astounding baseline for recognition.
\newblock In {\em Computer Vision and Pattern Recognition Workshops (CVPRW),
  2014 IEEE Conference on}, pp. 512--519. IEEE, 2014.

\bibitem{Razavian:wa}
A.~S. Razavian, H.~Azizpour, J.~Sullivan, and S.~Carlsson.
\newblock Cnn features off-the-shelf: an astounding baseline for recognition.
\newblock In {\em Computer Vision and Pattern Recognition Workshops (CVPRW),
  2014 IEEE Conference on}, pp. 512--519. IEEE, 2014.

\bibitem{Ribeiro:2016:WIT:2939672.2939778}
M.~T. Ribeiro, S.~Singh, and C.~Guestrin.
\newblock "why should i trust you?": Explaining the predictions of any
  classifier.
\newblock In {\em Proceedings of the 22Nd ACM SIGKDD International Conference
  on Knowledge Discovery and Data Mining}, KDD '16, pp. 1135--1144. ACM, New
  York, NY, USA, 2016. doi: {{%
10\hspace{.1pt}\discretionary{.}{%
}{.}\hspace{.4pt}1145\discretionary{/}{%
}{/}2939672\hspace{.1pt}\discretionary{.}{%
}{.}\hspace{.4pt}2939778}}


\bibitem{salimans2017pixelcnn++}
T.~Salimans, A.~Karpathy, X.~Chen, and D.~P. Kingma.
\newblock Pixelcnn++: Improving the pixelcnn with discretized logistic mixture
  likelihood and other modifications.
\newblock {\em arXiv preprint arXiv:1701.05517}, 2017.

\bibitem{schiffman1981introduction}
S.~S. Schiffman, M.~L. Reynolds, and F.~W. Young.
\newblock {\em Introduction to multidimensional scaling: Theory, methods, and
  applications}.
\newblock Emerald Group Publishing, 1981.

\bibitem{Zeitz:2018:BIV:3144687.3144715}
J.~Z. Self, M.~Dowling, J.~Wenskovitch, I.~Crandell, M.~Wang, L.~House,
  S.~Leman, and C.~North.
\newblock Observation-level and parametric interaction for high-dimensional
  data analysis.
\newblock {\em ACM Trans. Interact. Intell. Syst.}, 8(2):15:1--15:36, June
  2018. doi: {{%
10\hspace{.1pt}\discretionary{.}{%
}{.}\hspace{.4pt}1145\discretionary{/}{%
}{/}3158230}}


\bibitem{shipman1994supporting}
F.~M. Shipman~III and R.~McCall.
\newblock Supporting knowledge-base evolution with incremental formalization.
\newblock In {\em Proceedings of the SIGCHI Conference on Human Factors in
  Computing Systems}, pp. 285--291. ACM, 1994.

\bibitem{Smilkov:2017to}
D.~Smilkov, S.~Carter, D.~Sculley, F.~B. Vi{\'e}gas, and M.~Wattenberg.
\newblock {Direct-Manipulation Visualization of Deep Networks}.
\newblock {\em arxiv.org}, Aug. 2017.

\bibitem{WANG201766}
Y.~Wang, Z.~Luo, and P.-M. Jodoin.
\newblock Interactive deep learning method for segmenting moving objects.
\newblock {\em Pattern Recognition Letters}, 96:66 -- 75, 2017.
\newblock Scene Background Modeling and Initialization. doi: {{%
10\hspace{.1pt}\discretionary{.}{%
}{.}\hspace{.4pt}1016\discretionary{/}{%
}{/}j\hspace{.1pt}\discretionary{.}{%
}{.}\hspace{.4pt}patrec\hspace{.1pt}\discretionary{.}{%
}{.}\hspace{.4pt}2016\hspace{.1pt}\discretionary{.}{%
}{.}\hspace{.4pt}09\hspace{.1pt}\discretionary{.}{%
}{.}\hspace{.4pt}014}}


\bibitem{WOLD198737}
S.~Wold, K.~Esbensen, and P.~Geladi.
\newblock Principal component analysis.
\newblock {\em Chemometrics and Intelligent Laboratory Systems}, 2(1):37 -- 52,
  1987.
\newblock Proceedings of the Multivariate Statistical Workshop for Geologists
  and Geochemists. doi: {{%
10\hspace{.1pt}\discretionary{.}{%
}{.}\hspace{.4pt}1016\discretionary{/}{%
}{/}0169\discretionary{%
}{-}{-}7439\discretionary{%
}{(}{(}87\discretionary{)}{%
}{)}80084\discretionary{%
}{-}{-}9}}


\bibitem{1382935}
P.~C. Wong, B.~Hetzler, C.~Posse, M.~Whiting, S.~Havre, N.~Cramer, A.~Shah,
  M.~Singhal, A.~Turner, and J.~Thomas.
\newblock In-spire infovis 2004 contest entry.
\newblock In {\em IEEE Symposium on Information Visualization}, pp. r2--r2, Oct
  2004. doi: {{%
10\hspace{.1pt}\discretionary{.}{%
}{.}\hspace{.4pt}1109\discretionary{/}{%
}{/}INFVIS\hspace{.1pt}\discretionary{.}{%
}{.}\hspace{.4pt}2004\hspace{.1pt}\discretionary{.}{%
}{.}\hspace{.4pt}37}}


\bibitem{Wongsuphasawat:bb}
K.~Wongsuphasawat, D.~Smilkov, J.~Wexler, J.~Wilson, D.~Mané, D.~Fritz,
  D.~Krishnan, F.~B. Viégas, and M.~Wattenberg.
\newblock Visualizing dataflow graphs of deep learning models in tensorflow.
\newblock {\em IEEE Transactions on Visualization and Computer Graphics},
  24(1):1--12, Jan 2018. doi: {{%
10\hspace{.1pt}\discretionary{.}{%
}{.}\hspace{.4pt}1109\discretionary{/}{%
}{/}TVCG\hspace{.1pt}\discretionary{.}{%
}{.}\hspace{.4pt}2017\hspace{.1pt}\discretionary{.}{%
}{.}\hspace{.4pt}2744878}}


\bibitem{yang2007evaluating}
J.~Yang, Y.-G. Jiang, A.~G. Hauptmann, and C.-W. Ngo.
\newblock Evaluating bag-of-visual-words representations in scene
  classification.
\newblock In {\em Proceedings of the International Workshop on Workshop on
  Multimedia Information Retrieval}, MIR '07, pp. 197--206. ACM, New York, NY,
  USA, 2007. doi: {{%
10\hspace{.1pt}\discretionary{.}{%
}{.}\hspace{.4pt}1145\discretionary{/}{%
}{/}1290082\hspace{.1pt}\discretionary{.}{%
}{.}\hspace{.4pt}1290111}}


\bibitem{taskonomy2018}
A.~R. Zamir, A.~Sax, W.~B. Shen, L.~J. Guibas, J.~Malik, and S.~Savarese.
\newblock Taskonomy: Disentangling task transfer learning.
\newblock In {\em IEEE Conference on Computer Vision and Pattern Recognition
  (CVPR)}. IEEE, 2018.

\bibitem{Zeiler:2014fr}
M.~D. Zeiler and R.~Fergus.
\newblock {Visualizing and Understanding Convolutional Networks}.
\newblock In {\em Computer Vision {\textendash} ECCV 2014}, pp. 818--833.
  Springer, Cham, Cham, Sept. 2014.

\end{thebibliography}
\end{document}